\documentclass[12pt]{iopart}

\usepackage[utf8]{inputenc}
\usepackage{multirow,float} 
\usepackage{booktabs}    
\usepackage{subfigure} 
\usepackage{adjustbox} 
\usepackage{siunitx}
\usepackage{mathrsfs}
\usepackage{float}
\usepackage{amsfonts}
\usepackage{amssymb}
\usepackage{amsthm}

\expandafter\let\csname equation*\endcsname\relax
\expandafter\let\csname endequation*\endcsname\relax

\usepackage{amsmath}
\usepackage[dcucite,abbr]{harvard}
\usepackage{fancyhdr}	
\usepackage{graphicx}
\usepackage[section]{placeins}  
\usepackage[mathlines]{lineno}
\usepackage{tikz}
\usetikzlibrary{matrix,positioning}
\usepackage{placeins}
\usepackage{lineno,hyperref}
\usepackage[flushmargin,multiple]{footmisc} 
\usepackage{iopams}  
\usepackage{xcolor}
\usepackage{caption}

\begin{document}

\title[\centering MRI-driven Adaptive Proton Therapy: MRI-based DECT Synthesis using DDPM]
{MRI-Driven Dual-Energy CT Synthesis via Conditional Denoising Diffusion Probabilistic Models for MRI-only Radiotherapy and Protontherapy} 

\author{Juan Manuel Molina-Maza$^{1,2}$, Chih-Wei Chang$^{1}$, David Viar-Hernandez$^{1,2}$, Zach Eidex$^{1}$, Mojtaba Safari$^{1}$, Shaoyan Pan$^{1}$, Juan Maria Perez-Moreno$^{3}$, Norberto Malpica$^{2}$, Angel Torrado-Carvajal$^{2}$, Borja Rodriguez-Vila$^{2\diamond}$, Xiaofeng Yang$^{1\diamond}$}

\address{$^1$ Department of Radiation Oncology and Winship Cancer Institute, Emory University, Atlanta, USA}

\address{$^2$ Medical Image Analysis and Biometry Laboratory, Universidad Rey Juan Carlos, Madrid, Spain}

\address{$^3$ Departamento de Física Médica, Centro de Protonterapia Quironsalud, Madrid, Spain}

\address{$^\diamond$ Contributed equally to this work}

\vspace{10pt}
\begin{indented}
\item[] Corresponding authors: xiaofeng.yang@emory.edu and juan.molina.maza@emory.edu

\item[]\today
\end{indented}

\begin{abstract}

{\bf Objective:} Magnetic resonance imaging (MRI) provides excellent soft tissue contrast and avoids ionizing radiation, making it highly valuable in radiation therapy (RT) for tumor evaluation and localization, although not suitable for precise dose calculation. Dual-energy CT (DECT) scans enable superior tissue characterization and reduce range uncertainties over Single-Energy CT scans, but they expose the patient to ionizing radiation. We propose a denoising diffusion probabilistic model (DDPM) to generate synthetic high-energy (sHECT) and low-energy (sLECT) CT scans from a single MRI. \\

{\bf Approach:} Our head and neck MRI-DECT database was retrospectively collected from protontherapy (PT) patients and further preprocessed. A DDPM multidecoder framework was designed and trained to simultaneously generate sHECT and sLECT volumes from the MRI input. Two architectures (SwinVit and ResUNet) were used as the diffusion denoiser autoencoder, and they were compared against two conditional generative adversarial network (cGAN) approaches (Pix2PixGAN and CycleGAN). Synthesis fidelity and clinical utility of synthetic images were assessed via visual inspection, image quality metrics, Hounsfield Unit (HU) error distributions, anatomical profile, uncertainty level and stopping power ratio (SPR) maps. \\

{\bf Main Results:} Our DDPM outperformed cGANs in both visual similarity and image quality metrics: Mean Absolute Error (MAE) for sHECT and sLECT were $36.29 \pm 5.25$ and $43.72 \pm 6.77$ HU; Peak Signal-to-Noise Ratio (PSNR) for sHECT and sLECT were $27.44 \pm 1.06$ and $26.26 \pm 0.98$ dB; Structural Similarity Index Measure (SSIM) for sHECT and sLECT were $0.89 \pm 0.02$ and $0.87 \pm 0.03$; Normalized Cross-Correlation (NCC) for sHECT and sLECT were $0.95 \pm 0.01$ and $0.94 \pm 0.01$. Detailed structures and anatomic boundaries were recovered from sHECT and sLECT, while the uncertainty of their derived SPR maps stayed within the expected acceptable range. \\

{\bf Significance:} A novel DDPM approach was proposed for synthesizing DECT images from MRI scans in a cohort of PT patients. We spare patients from the harmful radiation exposure of DECT, while we obtain the high-quality material-specific information, SPR robust estimation and proton-range accuracy required in PT workflow.

\end{abstract}




\setlength{\baselineskip}{0.7cm}      

\pagenumbering{arabic}
\setcounter{page}{1}
\pagestyle{fancy}

\section{Introduction}

{Precise imaging is essential in radiation therapy (RT) for accurate dose delivery, minimal radiation exposure of the adjacent healthy tissues and organs-at-risk (OAR), and effective treatment planning \cite{pereira2014role}. In modern RT workflows, treatment planning systems rely on Computed Tomography (CT) scans to directly quantify patient-specific material properties and ensure reliable dose deposition. The accuracy of this material quantification is far more critical in proton therapy (PT), where even small uncertainties in SPR estimation translate into shifts on the Bragg peak position that may lead to under-dosage of the target or over-dosage of nearby OAR \cite{lane2023image}. The uncertainty in CT-to-SPR estimation represents the dominant source of proton range uncertainty in treatment planning scenarios, and forces clinicians to apply larger safety margins around clinical target volumes, exposing surrounding healthy tissues to unnecessary radiation doses and diminishing the potential of PT treatment \cite{paganetti2012range}.}

As opposed to the limitations of single-energy CT (SECT) spectrum for differentiating subtle tissue properties, where multiple tissue compositions can exhibit identical Hounsfield Units (HU) and lead to inaccuracies in SPR calculations \cite{bar2017potential}, dual-energy CT (DECT) acquisition protocols employ two different X-ray energy spectra yielding high-energy CT (HECT) and low-energy CT (LECT) images. They provide superior spectral tissue characterization due to the unique energy-dependent attenuation profiles of different materials {\cite{peters2022vivo,wohlfahrt2020refinement}}, facilitating accurate differentiation and decomposition of tissues as well as improved imaging contrast \cite{rutherford1976x,mccollough2020principles}. Since DECT significantly reduces uncertainties in proton range calculations \cite{peters2022reduction}, clinicians can enhance the targeting precision to confidently reduce treatment margins. The dose released to the tumor is therefore intensified while we significantly spare the damage to surrounding healthy tissues and OAR, thus improving the balance between treatment effectiveness and patient safety, which ultimately leads to better clinical outcomes \cite{wohlfahrt2017dual,taasti2023clinical}.

Magnetic Resonance Imaging (MRI) also represents an integral component in RT due to the superior soft tissue contrast and precise tumor and OAR delineation \cite{chandarana2018emerging}, while preventing patient exposure to harmful radiation. This makes MRI-only workflows a promising alternative in RT, also avoiding the uncertainties associated with multimodal imaging registration {\cite{ulin2010results}}, and reducing the extra expenditure, additional time investments and increased procedural costs that arise whenever a patient must undergo both CT and MRI acquisitions instead of just running the MRI session {\cite{persson2023mri}}. Nevertheless, since MRI acquisitions fundamentally depend on hydrogen atom responses rather than comprehensive material composition, they cannot directly provide clinically relevant material properties in RT like electron density and SPR \cite{hoffmann2020mr}. 

Several approaches have been proposed to address this MRI constraint. Segmentation-based approaches classify tissues into discrete anatomical categories by assigning a predefined CT attenuation value to each category, but they are vulnerable to classification errors and unable to represent the detailed structure of tissues \cite{kristensen2008dosimetric,hsu2013investigation}. Atlas-based frameworks employ previously acquired CT images as references, but the requirement of precise anatomical alignment poses challenges due to inter-patient variability and deformation \cite{uh2014mri,sjolund2015generating}. Deep learning (DL) techniques have shown promise in generating high-quality synthetic CT (sCT) images from MRI data for radiation oncology applications \cite{eidex2024deep,boulanger2021deep,sherwani2024systematic}. {Other studies have shown the potential of DL frameworks to estimate mass density and SPR maps using MRI \cite{gao2024mri}, DECT \cite{chang2022dual} or both imaging modalities \cite{chang2023multimodal}.}  

{The considerations presented along this section highlight the need to join the gap between MRI-only RT and accurate material tissue characterization. MRI-only workflows represent a highly attractive alternative in RT, but their clinical translation is hindered by the inability of MRI to compute radiological properties necessary for RT treatment planning. This limitation becomes especially relevant in PT, where CT-to-SPR estimation dominates the overall proton-range uncertainty. DECT has become an increasingly important component of modern PT workflows demonstrating the enhancement in SPR accuracy for more reliable dose calculation and safer treatment. In this context, synthesizing DECT images from MRI represents a substantial step forward in RT practice. It enables the use of MRI-only workflows to avoid the harmful DECT radiation exposure, while obtaining the high-quality material-specific information, SPR robust estimation and proton-range accuracy required in PT that are currently achievable only through DECT.}

{The present work addresses this unmet need by leveraging a conditional denoising diffusion probabilistic model (DDPM) to generate synthetic DECT (sDECT) images from MRI volumes, employing a retrospective patient database. By modeling the diffusion process as a sequential denoising procedure, Gaussian noise was iteratively converted into realistic images, avoiding the training limitations and instability of conditional generative adversarial networks (cGANs). The evaluation of clinical realism and quantitative accuracy of sDECT images revealed improved image quality, and showed the effective representation of intricate anatomical regions, complex structures and subtle differences between low- and high-energy levels. A systematic comparison was also performed to measure the error and uncertainty of SPR maps derived from our sDECT images with respect to those obtained from GT DECT acquisitions. Quantitative error metrics, population-level distributional analysis and voxel-wise error maps obtained for sDECT images and their SPR maps revealed promising results, showing the potential of this work to streamline and improve the current PT workflow.}

\section{Related Work}

Diverse DL approaches including Convolutional Neural Networks (CNNs), Vision Transformers, cGANs, and diffusion models have been explored to generate MRI-based sCT and sDECT images in radiation oncology. Most of these methods focus on synthesizing SECT images. Early investigations demonstrated promising results employing diverse Dixon images for human pelvis CT generation with U-Net architectures \cite{bahrami2020new}. Two-dimensional cGANs have been extensively implemented across different clinical scenarios \cite{qi2020multi,maspero2018dose,xu2019multichannel,tie2020pseudo,koike2020feasibility,maspero2020deep}. Experimental studies have leveraged different MRI sequences, including Dixon water, fat, and in-phase pelvis MRI \cite{maspero2018dose}, and multi-sequence abdominal acquisitions \cite{xu2019multichannel}. Comparative analyses revealed that integrating multiple MRI sequences consistently outperformed single-sequence approaches \cite{qi2020multi}, particularly T1-weighted (T1-w) and T2-weighted (T2-w). Specific clinical applications have included nasopharyngeal carcinoma studies using T1-w and T2-w head acquisitions \cite{tie2020pseudo}, brain RT investigations incorporating T1-w, T2-w, and T2-FLAIR MRI sequences \cite{koike2020feasibility}, and pediatric brain dose calculation evaluations trained on single T1-w MRI datasets \cite{maspero2020deep}. CycleGANs have also been utilized in RT for modality translation in various anatomical regions, including brain and abdomen MRI \cite{yang2020unsupervised}, pelvic T2-w MRI \cite{brou2021improving}, and head and neck imaging using a multi-CycleGAN approach \cite{liu2021ct}. DDPMs have also been employed for MR-to-CT image synthesis task, utilizing male pelvic T2-w MRI \cite{lyu2022conversion}, T1-w and T2-w MRI cervical, thoracic, and lumbar \cite{graf2023denoising}, and brain and prostate T1-w MRI \cite{pan2024synthetic}. Considering DECT image synthesis, some works have developed DDPM approaches using imaging modalities like CBCT \cite{viar2024exploring} or SECT \cite{gao2024ct}.

We have compiled throughout this section several DL approaches for MRI-based sCT and sDECT generation in radiation oncology. Notably, only the work of \cite{liu2021synthetic} generated a sDECT volume from MRI data, by means of a label-GAN. To our knowledge, no diffusion-based approaches focused on generating MRI-based sDECT images have yet been reported in the literature. 

\section{Dataset}

This retrospective study received institutional review board approval with a waiver of informed consent while maintaining HIPAA compliance. The cohort comprised 53 patients diagnosed with head and neck cancer (19 women, age range 21-89 years old). All patients underwent same-day imaging with both DECT and MRI acquisitions as part of their PT planning protocol. DECT scans were performed using a Siemens SOMATOM Definition Edge Twin-beam scanner (Siemens Healthineers, Erlangen, Germany) with patients in treatment position, and they were processed with Siemens Syngo CT VA48A software. Acquisition parameters included: 120 kVp, 0.45 pitch, 500 ms of exposure time, 0.5-second rotation time, approximately 30-second scan duration, CTDIvol of approximately 20 mGy, tube current between 500-650 mA, matrix size of 512 × 512, and 0.98 × 0.98 × 1 mm³ of spacing. Images were reconstructed using the SAFIRE algorithm with Q30f kernel and metal artifact reduction. The system utilized 0.6 mm tin and 0.05 mm gold filtration to split the 120 kVp beam into high-energy and low-energy components for separate reconstructions. MRI acquisition utilized a Siemens Aera 1.5T scanner to generate 3D T1-weighted sequences with the following parameters: TE/TR = 2.4/7.8 ms; 12° flip angle; 1.2 × 1.2 × 1.2 mm³ isotropic resolution; echo time = 2.39 ms; matrix size = 512 × 512. {The MRI volume of every patient went through deformable registration and further resampling to match DECT geometry and position. We utilized Velocity AI 3.2 (Varian Medical Systems, Palo Alto, CA), a commercially approved and widely adopted deformable image registration platform that has been extensively validated in clinical RT workflows, to register MRI to both LECT and HECT volumes}. All deformation results were independently reviewed by two medical physicists to ensure the accuracy and quality of the registration. The original image volumes in DICOM format were converted to NIFTI format. Signal intensity standardization was then implemented through a min-max normalization procedure, which rescaled the pixel values to a normalized range $[0,1]$ for the Diffusion methods, and $[-1,1]$ for the cGAN methods. Finally, original image matrix size was cropped from 512x512 to 256x256 dimensions.

\section{Methods}

\subsection{Denoising Diffusion Probabilistic Model}

\begin{figure*}[!h]
    \centering
    \includegraphics[width=\textwidth]{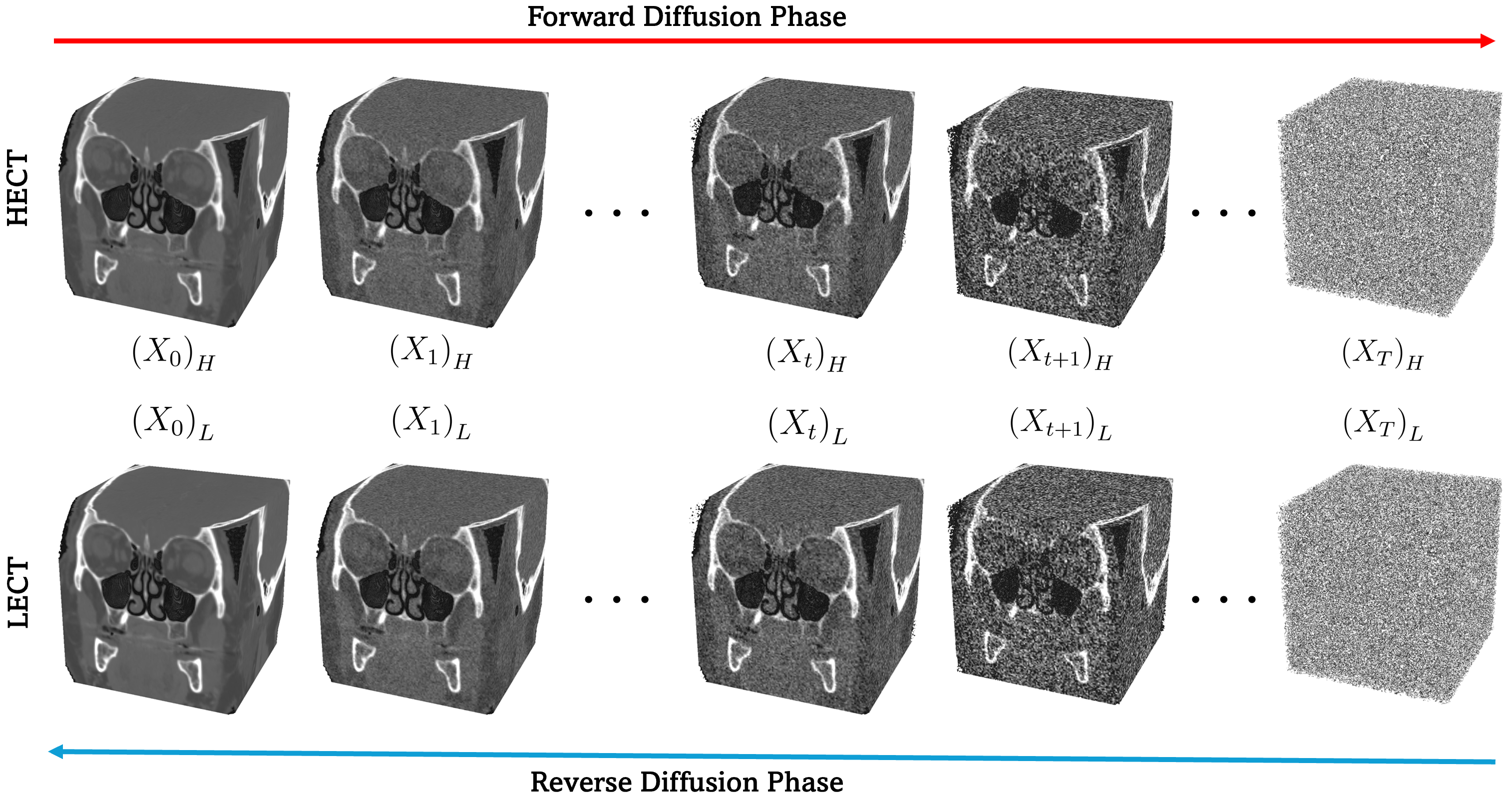}
    \caption{Visual representation of the diffusion process for both HECT (upper row) and LECT (lower row) volumes, showing the forward (red arrow) and reverse (blue arrow) diffusion phases. In the forward phase, a small amount of noise is iteratively added to each original volume until only Gaussian noise remains. During the reverse phase, our neural network progressively removes the noise at each timestep, learning to recover high-fidelity 3D volumes for both energy levels.}
    \label{fig:diffusion_process}
\end{figure*}

Based on previous diffusion-based works \cite{ho2020denoising,dhariwal2021diffusion}, we developed a DDPM framework that converts an MRI scan into an sDECT image. A 3D isotropic Gaussian noise volume $V\sim\mathcal{N}(0,I)$ is transformed into an sDECT image, the latter being conditioned on the corresponding MRI data. By adding a small random noise component $\epsilon$ to the ground truth (GT) DECT image $X$ through $T$ successive timesteps, it is possible to transform $X$ into a purely Gaussian noise sample $V$. When $T$ is large enough, the original noise-free image $X$ can be reconstructed by systematically reversing this process and removing the added noise from $V$. In essence, our DDPM framework process consists of two consecutive phases: Forward Diffusion and Reverse Diffusion (Figure \ref{fig:diffusion_process}). The forward phase injects Gaussian noise into a DECT image so that it becomes entirely noise-driven, while the reverse phase (learned by our DL neural network) removes the noise and retrieves the sDECT image.

\subsubsection{Forward Diffusion Phase\\} 

This forward diffusion procedure is modeled as a Markov chain $q$, where each noisy image at time step \(t\) depends only on the noisy image at the previous time step \(t-1\). The probability of shifting from \({X}_{t-1}\) to \({X}_{t}\) follows a Gaussian distribution $N$:

\begin{equation}
\label{eq1forwarddiffusion}
q\bigl(X_t \mid X_{t-1}\bigr)_{L,H}
\,=\,
\mathcal{N}\!\Bigl(
X_t;\,
\sqrt{1 - \beta_t}\,X_{t-1},\;
\beta_t\,I
\Bigr)
\end{equation}

where $\beta_t$ represents the pre-determined variance at $t$ timestep. Moreover, the noisy images at any discretional $t$ timestep can be defined as follows:

\begin{equation}
\label{eq2forwarddiffusion}
X_t
=
\sqrt{\prod_{i=1}^T \bigl(1 - \beta_i\bigr)} \, X_0
\;+\;
\sqrt{1 - \prod_{i=1}^T \bigl(1 - \beta_i\bigr)} \, \epsilon_t
\end{equation}

where $\epsilon_{t} \sim\mathcal{N}(0,I)$. Starting with LECT \(\bigl({X}_{0})_{L}\) and HECT \(\bigl({X}_{0})_{H}\) GT scans, noise is iteratively added across \(t\) time steps, eventually converting the original DECT images into Gaussian noise \({X}_{T}\). Timestep \(t\) is randomly chosen during the training phase within a range from 1 to 1000 in increments of 50. Noise is independently added to \(\bigl({X}_{0})_{L}\) and \(\bigl({X}_{0})_{H}\) at any chosen $t$ timestep, where the resulting noisy images act as the input to the diffusion model, aiming to estimate both the mean and the variance of the Gaussian noise added at that specific step. 

\subsubsection{Reverse Diffusion Phase\\}

Along the reverse diffusion process, the denoised version $\left(\hat{X}_0 \right)_{L,H}$ of DECT original images is gradually recovered by systematically cleaning the noise that was introduced in the previous forward diffusion stage. This denoising procedure is driven by training DL methods, which predict the image at timestep $t-1$ from the noisier version at timestep $t$. Once an image $\left({X}_{T} \right)_{L,H}$ has accumulated enough noise, it can be eventually recovered as its noise-free state version $\left(\hat{X}_0 \right)_{L,H}$. Due to the random nature of the backward diffusion steps, the resulting sDECT images can sometimes diverge from the actual patient data. To resolve this ambiguity, a conditional MRI volume is therefore incorporated into the process. This ensures that the sDECT images remain fully aligned to the patient's specific anatomy and structure, so that each sDECT output directly corresponds to the given MRI. This reverse diffusion process is formulated as follows:

\begin{equation}
\label{eq1reversediffusion}
p\bigl(X_{t-1} \mid X_{t}\bigr)_{L,H}
=
\mathcal{N}\bigl(X_{t-1};\,\mu_{t},\,\Sigma_{t}\bigr)
\end{equation}

where $\mu_{t}$ and $\Sigma_{t}$ represent the mean and covariance matrix for every $t$ timestep, and $N$ is the Gaussian distribution used to model this process. However, it is only possible to directly compute this inverse probability when the noise-free DECT image $\left({X}_{0} \right)_{L,H}$ is already available, but $\left({X}_{0} \right)_{L,H}$ is not known for test data during inference phase. As this direct calculation becomes infeasible, we use $\theta$ learned parameters from our DL neural network to approximate the ideal probability distribution $p\bigl(X_{t-1} \mid X_{t}\bigr)_{L,H}$ as follows:

\begin{equation}
\label{eq2reversediffusion}
p_{\theta}\bigl(X_{t-1} \mid X_{t}, Z\bigr)_{L,H}
=
\mathcal{N}\Bigl(
X_{t-1};\,
\mu_{\theta}\bigl(X_{t}, t \mid Z\bigr),\,
\Sigma_{\theta}\bigl(X_{t}, t \mid Z\bigr)
\Bigr)
\end{equation}

where $Z$ is the conditioning MRI data, $\mu_{\theta}$ is the estimated mean matrix, and $\Sigma_{\theta}$ is the estimated variance matrix of the inverse Gaussian distribution. Therefore, we formulate the less noisy image $X_{t-1}$ as follows:

\begin{equation}
\label{eq3reversediffusion}
X_{t-1}
=
\mu_{\theta}\bigl(X_{t}, t \mid Z\bigr)
\;+\;
\sigma_{\theta}\bigl(X_{t}, t \mid Z\bigr)
\,*\epsilon
\end{equation}

where $\epsilon \sim\mathcal{N}(0,I)$ refers to the noise generated from the Gaussian distribution, and $\sigma_{\theta}$ is the standard deviation of the reverse distribution.  

\subsubsection{Technical implementation \\}

The reverse diffusion process produces two key outputs: the mean matrix $\mu_{\theta}$ and the covariance matrix $\Sigma_{\theta}$. However, directly predicting the mean at $X_{t-1}$ often yields unstable training outcomes, partly because there is no explicit constraint on the mean’s value \cite{ho2020denoising}. To improve both the stability and the reliability of the sDECT outputs, the DL model is instead trained to predict the mean of the added noise epsilon. By focusing on noise prediction, error propagation is reduced, and we can ultimately reconstruct higher-fidelity sDECT images. The mean $\mu_{\theta}$ and the estimated noise $\epsilon_{\theta}$ are both related by the following equation:

\begin{equation}
\mu_\theta\bigl(X_t, t \mid Z\bigr)
= \frac{1}{\sqrt{1 - \beta_t}}
  \left(
    X_t \;-\;
    \frac{\beta_t}{
      \sqrt{1 - \sum_{i=1}^t \bigl(1 - \beta_i\bigr)}
    }
    \,\epsilon_\theta\!\bigl(X_t, t \mid Z\bigr)
  \right)
\end{equation}

Throughout the training phase, the objective is to accurately model the noise that is introduced at each discrete $t$ time step. In practice, this optimization involves minimizing the Mean Absolute Error (MAE) between the predicted noise $\epsilon_{\theta}$ at a given time step $t$ and the corresponding GT noise $\epsilon_{t}$ across all possible $t$ timesteps:

\begin{equation}
\label{eq1technicalimplementation}
L_{\mu_{\theta}} = \operatorname*{argmin}_{\theta} \mathrm{MAE}\ \bigl(\epsilon_{t}, \epsilon_{\theta}(X_{t}, t \mid Z)\bigr)
\end{equation}

The variance coefficient $k_{\theta}(x_t,t|Z)$ is first predicted by the DL neural network at every $t$ timestep in order to compute afterwards the covariance matrix $\Sigma_{\theta}$ \cite{dhariwal2021diffusion}. Both parameters are related by the following equation:    

\begin{equation}
\label{eq2technicalimplementation}
\Sigma_{\theta}(x_t,t|Z)_{L,H} = \exp \left( \frac{k_{\theta}(x_t,t|Z) + 1}{2} \cdot \log \beta_t + \frac{1 - k_{\theta}(x_t,t|Z)}{2} \cdot \log \left( \frac{\prod_{i=1}^{t-1}(1 - \beta_i)}{1 - \prod_{i=1}^{t}(1 - \beta_i)} \right) \right)
\end{equation}

Based on previous approaches \cite{dhariwal2021diffusion}, a variational lower bound (VLB) loss serves to optimize the variance:

\begin{equation}
\label{eq3technicalimplementation}
L_{\Sigma_{\theta}} 
= \operatorname*{argmin}_{\theta} 
  \mathrm{VLB}\ \Bigl(
    N\bigl(X_{t-1}; \mu_{t}, \Sigma_{t}\bigr),
    N\bigl(X_{t-1}; \mu_{t}, \Sigma_{t}^{\theta}\bigr)
  \Bigr)
\end{equation}

Finally, we can optimize our DL neural network parameters by minimizing the following main loss function:

\begin{equation}
\label{eq4technicalimplementation}
L = L_{\mu_{{\theta}}} + \gamma\,L_{\Sigma_{{\theta}}}
\end{equation}

where $\gamma$ is chosen using an empirical process which is guided by the proportion of the diffusion steps during inference relative to those used in training.

In summary, our DL architecture processes the noisy image input $X_{t}$ from random $t$ timestep, conditioned by an image $Z$ (MRI scan). By estimating the mean $\mu_{\theta}$ and standard deviation $\sigma_{\theta}$ of the reverse probability distribution along training phase, a cleaner image $\hat{X}_0$ is reconstructed during inference phase all without needing any prior knowledge of the original clean image $X_{0}$.

\subsection{Neural Network Architecture}

Before the training phase begins, noise is independently applied to LECT and HECT images using a randomly selected $t$ timestep, which produces separate mean and variance matrices for low- and high-energy levels, respectively. By handling them separately, the model can tackle the particular noise patterns associated with every energy level, achieving a more precise DECT image reconstruction. Nevertheless, DECT images are unavailable during the inference phase so instead they are replaced with noise channels, each of them being concatenated with the MRI volume to constitute the input to the neural network.

We employed two distinct three-dimensional neural network architectures to generate sDECT images with our DDPM: SwinVit and ResUnet. The primary distinction between the two versions lies in the underlying structural design of the backbone: whereas ResUNet is built upon convolutional blocks to preserve the local spatial information and capture the specific local features related to each volume, SwinVit combines these convolutional blocks with Swin Transformer (ST) blocks. By incorporating shifted windows alongside self-attention operations, ST architecture model long-range dependencies and effectively capture both fine-grained tissue details and global texture features, an advantage that proves especially valuable in robust feature integration between both energy levels. Residual connections were applied within both ST blocks and convolutional blocks to maintain feature continuity. Skip connections were also used in both ResUnet and SwinVit architectures to link each encoder-decoder counterpart by concatenating their corresponding feature maps to maintain spatial information among different feature map complexities. Every convolutional block contains a pair of convolutional layers (kernel size = 3x3x3), each of them followed by a batch normalization layer and a SiLu activation function. 

The original version of the ST block \cite{liu2021swin} was modified to process 3D medical image volumes. The ST block comprises two key backbone components: a self-attention window (W-SA) module and a self-attention shifted-window (SW-SA) module. The SW-SA module is structurally identical to the W-SA module, but the latter shifts the non-overlapping windows by half of the window size, enabling attention across window boundaries and improving the model’s capacity to capture long-range dependencies throughout the image. SW-SA module is applied right after W-SA module. For an input feature map $F \in \mathbb{R}^{H \times W \times L \times C}$, non-overlapping windows are partitioned by the W-SA layer (number of windows $= \frac{H}{N} \times \frac{W}{N} \times \frac{L}{N_L} \times C$, window size $= (N/2, N/2, N_{L}/2)$). After that, a multi-head self-attention (MHSA) layer with $P$ parallel heads applied independently within each window. The resulting head attention maps are concatenated and are finally reassembled to match the initial feature map size. The attention map within each head is computed as follows:

\begin{equation}
    h_p = \text{Softmax}\left(\frac{Q_pK_p^\top}{\sqrt{d_k}}\right)V_p
\end{equation}

Here, $Q_{p}$, $K_{p}$ and $V_{p}$ denote the independent Query, Key and Value weight matrices, respectively, for the head $h_p$. ST hyperparameters are the following: window size = $(4,4,4)$, patch size = $(4,4,4)$, initial embedding dimension = $256$, MLP size = $512$, number of heads = $8$. If one wishes to explore the technical details of ST more deeply, please refer to the original work \cite{liu2021swin}. 

\begin{figure*}[!h]
    \centering
    \includegraphics[width=\textwidth]{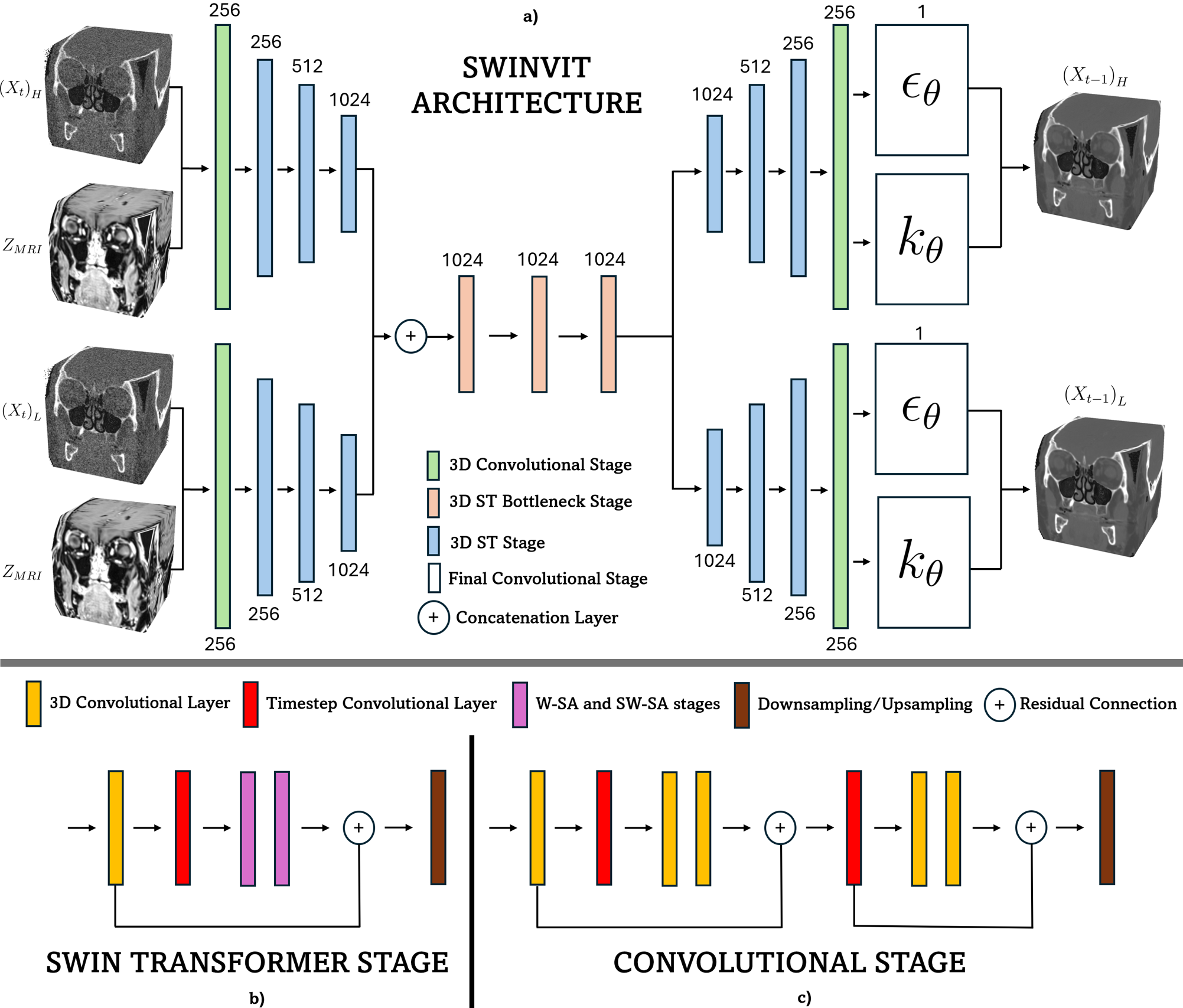}
    \caption{(a) Overview of the proposed SwinVit-based network for MRI-to-DECT synthesis during the reverse diffusion process: the model takes the current noisy HECT/LECT volumes as input alongside an MRI reference, then processes them through a 3D encoder-decoder pipeline with a timestep embedding applied to every stage of the network, yielding predictions for the noise and variance, which refine the HECT/LECT volumes. b) Schematic of the 3D ST stage and c) 3D Convolutional stage. ResUnet has a similar structure compared to SwinVit, but changes ST stages to Convolutional Stages.}
    \label{fig:neural_network_architecture}
\end{figure*}

The overall neural network structure relies on two parallel encoders with four encoding stages, a bottleneck with three stages, and two parallel decoders with four decoding stages (Figure \ref{fig:neural_network_architecture}). Parallel encoders allow the network to initially process each energy level input independently, merging their representations only at the end of the encoding phase to enable joint feature learning from the latent space. Prior to entering the bottleneck, feature maps from both encoder branches are concatenated to form a unified latent representation of the two energy levels. The bottleneck itself consists of three sequential stages without any downsampling or upsampling, designed to enhance global feature interactions and strengthen contextual understanding across the extracted features. Two parallel decoders predict the noise present in LECT and HECT images respectively, restoring the original shape of DECT images by taking the latent feature maps derived from the bottleneck.

In ResUnet architecture, each stage contains two convolutional blocks. The encoding stage reduces the image size with pooling, thereby reducing the complexity of feature maps, while the decoding stage returns the feature maps to the original image dimensions. The initial number of filters is 64 and is doubled/halved for encoding/decoding stages, where max-pooling/transposed convolutions operations are used for downsampling/upsampling. On the other hand, each stage of SwinVit architecture has one ST block, except for the first encoding stage and the last decoding stage, which contain two convolutional blocks. For every encoder/decoder stage, a patch merging/patch-expanding layer reduces/increases the number of patches by half/double and duplicates/halves the channel dimension. In the case of decoding stages, patch embedding layers keep the consistency in the number of channels by halving them after the skip connections.

Finally, the output of each decoder is fed into a final convolutional layer with 2 channels, which generates two different maps for each energy level: the added noise $\epsilon_{\theta}$ and its variance $k_{\theta}$. These maps are then used to compute the denoised clean version of DECT images at the $t-1$ timestep.

Moreover, for managing the progression of the diffusion process within DDPM, the representation of time-specific information is enabled by means of time embeddings using a sinusoidal embedding technique with a feature dimension of 128 and a maximum period of $10^6$. By incorporating this time-aware mechanism, the model can recognize and interpret evolving temporal patterns. This ensures accurate noise reduction without sacrificing meaningful data details and promotes the generation of clearer and high-quality reliable images. To align these temporal embeddings with the dimensions required by subsequent layers, an embedding expansion layer is employed. A given input timestep embedding $T \in \mathbb{R}^{1 \times d}$ is processed by this timestep expanding layer to output a scale feature map $sc \in \mathbb{R}^{1 \times d_c}$ and a shift feature map $sh \in \mathbb{R}^{1 \times d_c}$, where $d$ is the timestep dimension and $d_c$ is the feature map dimension. Finally, these expanded embedding maps ($sc$ and $sh$) are integrated with the outputs from layer $h$ by using scale-shift normalization \cite{dhariwal2021diffusion} as follows:

\begin{equation}
\label{eq:timeembedding}
O = h \ast (1 + sc) + sh
\end{equation}

All layers of the model employ this accurate temporal guidance to help enhance both denoising performance and the fidelity of sDECT images generated during the training phase.

\subsection{{Evaluation}}

\subsubsection{{Synthetic Image Evaluation} \\}

{The goal of this work is to synthesize two output images (HECT and LECT) from a MRI input. For performance assessment, we selected four metrics that quantify the similarity or discrepancy between sDECT and GT DECT images: Mean Absolute Error (MAE), Peak Signal-to-Noise Ratio (PSNR), Structural Similarity Index (SSIM), and Normalized Cross-Correlation (NCC).}

{MAE measures the average magnitude of the errors between two images by computing the mean of the absolute voxel-wise intensity differences and it is expressed in HU:}

\begin{equation}
{MAE (\hat{\textbf{y}}, \textbf{y}) =  \frac{1}{N}\sum_{i=1}^{N} | \hat{\textbf{y}}(i) - \textbf{y}(i) |} 
\label{MAEequation}
\end{equation}

{PSNR measures the extent to which noise degrades image quality, reporting the metric value on a logarithmic scale in decibels (dB):}

\begin{equation}
{PSNR (\hat{\textbf{y}}, \textbf{y}) =  20\log_{10} \left(\frac{MAX}{\sqrt{\frac{1}{N}\sum_{i=1}^{N} [ \hat{\textbf{y}}(i) - \textbf{y}(i) ]^{2}}}\right)} 
\label{PSNRequation}
\end{equation}

{where $MAX$ represents the maximum intensity value of the GT image.}

{SSIM measures image degradation by comparing three perceptual components: luminance, contrast, and structure. The values range from -1 to 1, with 1 indicating identical images. This dimensionless metric represents a quantitative approximation of human visual perception:}

\begin{equation}
{SSIM (\hat{\textbf{y}}, \textbf{y}) =  \frac{ (2 \mu_{ \hat{\textbf{y}} } \mu_{\textbf{y}} + c_{1})
                                       (2 \sigma_{\hat{\textbf{y}} \textbf{y}} + c_{2}) }
                                       { (\mu_{ \hat{\textbf{y}} }^{2} + \mu_{ \textbf{y} }^{2} + c_{1})
                                       (\sigma_{ \hat{\textbf{y}} }^{2} + \sigma_{ \textbf{y} }^{2} + c_{2})}} 
\label{SSIMequation}
\end{equation}

{where $c_{1}$ and $c_{2}$ are regularization constants of the metric itself, $ \sigma_{\hat{\textbf{y}} \textbf{y}}$ is the images joint covariance, $ \mu_{ \hat{\textbf{y}} } $ and $ \mu_{\textbf{y}} $ are the averages of images, and $ \sigma_{ \hat{\textbf{y}} }^{2} $ y $ \sigma_{ \textbf{y} }^{2} $ are the images variances.}

{NCC measures the structural-consistency of images by quantifying their correlation. Due to its insensitiveness to linear shifts in brightness and contrast, NCC ensures stability under varying illumination levels:}

\begin{equation}
{NCC (\hat{\textbf{y}}, \textbf{y}) = \frac{1}{N} \sum_{i=1}^{N} 
\frac{(\textbf{y}(i) - \mu_{\textbf{y}})(\hat{\textbf{y}}(i) - \mu_{\hat{\textbf{y}}})}
{\sigma_{\textbf{y}}\, \sigma_{\hat{\textbf{y}}}}}
\label{NCCequation}
\end{equation}

{where $ \mu_{\hat{\textbf{y}}} $ and $\mu_{\textbf{y}}$ are the sDECT and DECT image average, $\sigma_{\hat{\textbf{y}}}$ and $\sigma_{\textbf{y}}$ are the sDECT and DECT image standard deviation.}

\subsubsection{{SPR Evaluation} \\}

{A physics-based stoichiometric dual-energy method was used to generate the SPR maps, modeling the effective atomic number as a function of the HU ratio between LECT and HECT images by using empirically calibrated parameters. These parameters are obtained by scanning a phantom containing multiple material rods with known compositions and effective atomic numbers, and measuring their CT-number ratios at both high and low energies. Once calibrated, the model estimates the effective atomic number of unknown materials from their DECT-number ratios, and the resulting value is converted to SPR using the Bethe–Bloch equation. The Gammex RMI 467 electron-density phantom was employed, whose manufacturer provides the chemical compositions of the inserts. Although the phantom exceeds the size of a human head, its dimensions allow inclusion of a sufficient number of rods. SPR values for each rod were computed from their known compositions following \cite{schneider1996calibration}. Rods were randomly repositioned across five repeated scans to reduce bias from spatially correlated artifacts, and the same CT protocol was applied for phantom and patient imaging. HU values for all rods were measured in both energy channels and a calibration based on the SPR values calculated in \cite{su2018machine} was finally applied to the patient DECT images to obtain the SPR values.}

{SPR maps were calculated from sDECT and GT DECT images over test subset for each of the four DL-based models. The accuracy of the sDECT-derived SPR maps was quantified through four different metrics: Normalized Mean Absolute Error (NMAE), Normalized Mean Error (NME), PSNR and NCC. NMAE measures the absolute relative error while NME measures the relative error. They are formulated as follows:}

\begin{equation}
{NMAE = \frac{1}{N} \sum_{i=1}^{N} 
\frac{\left| \textbf{SPR}_{DECT}(i) - \textbf{SPR}_{sDECT}(i) \right|}{\textbf{SPR}_{DECT}(i)}}
\end{equation}

\begin{equation}
{NME = \frac{1}{N} \sum_{i=1}^{N} 
\frac{\left( \textbf{SPR}_{DECT}(i) - \textbf{SPR}_{sDECT}(i) \right)}{\textbf{SPR}_{DECT}(i)}}
\end{equation}

\subsection{Experiments}

The patient cohort was randomly allocated in three subsets, with approximately 70\% assigned to the training set, 10\% to the validation set, and 20\% to the test set, ensuring each patient appears exclusively in a single subset. Network parameters were randomly initialized, and optimization was conducted using Adam optimizer (learning rate: 0.00001, weight decay: 0.0001) across 200 epochs. Our comprehensive evaluation included multiple sDECT generation experiments: Diffusion-SwinVit, Diffusion-ResUnet, Pix2PixGAN, and CycleGAN. Because neither of our two cGAN architectures are supposed to produce HECT and LECT images in a single pass, each model was trained twice: once to generate HECT images, and again to generate LECT images. For our diffusion-based approach, MAE served as the loss function, calculated per batch exclusively on DECT tissue regions while excluding background pixels. We implemented selective parameter freezing, with one of the two decoders' parameters remaining fixed during optimization of the other decoder. This strategy ensured that each decoder (HECT or LECT) was updated based solely on its specific error metric, preserving independent latent space representations without cross-contamination or interference between the two energy CT types. All neural networks were implemented using PyTorch and trained on an NVIDIA A100 80GB GPU. Memory limitations forced us to divide the original 3D volumes into 64×64×64 voxel cubic patches ($12\,691$ unique patches for training). To prevent visible boundaries in the final reconstructed volume, patches were extracted using a half-overlap strategy. Following the prediction of the sDECT patches, overlapping regions were cropped to obtain patches of 32×32×32, which were then seamlessly combined to generate the complete sDECT volumes. 

\section{Results} 

\begin{figure*}[!h]
    \centering
    \includegraphics[width=\textwidth]{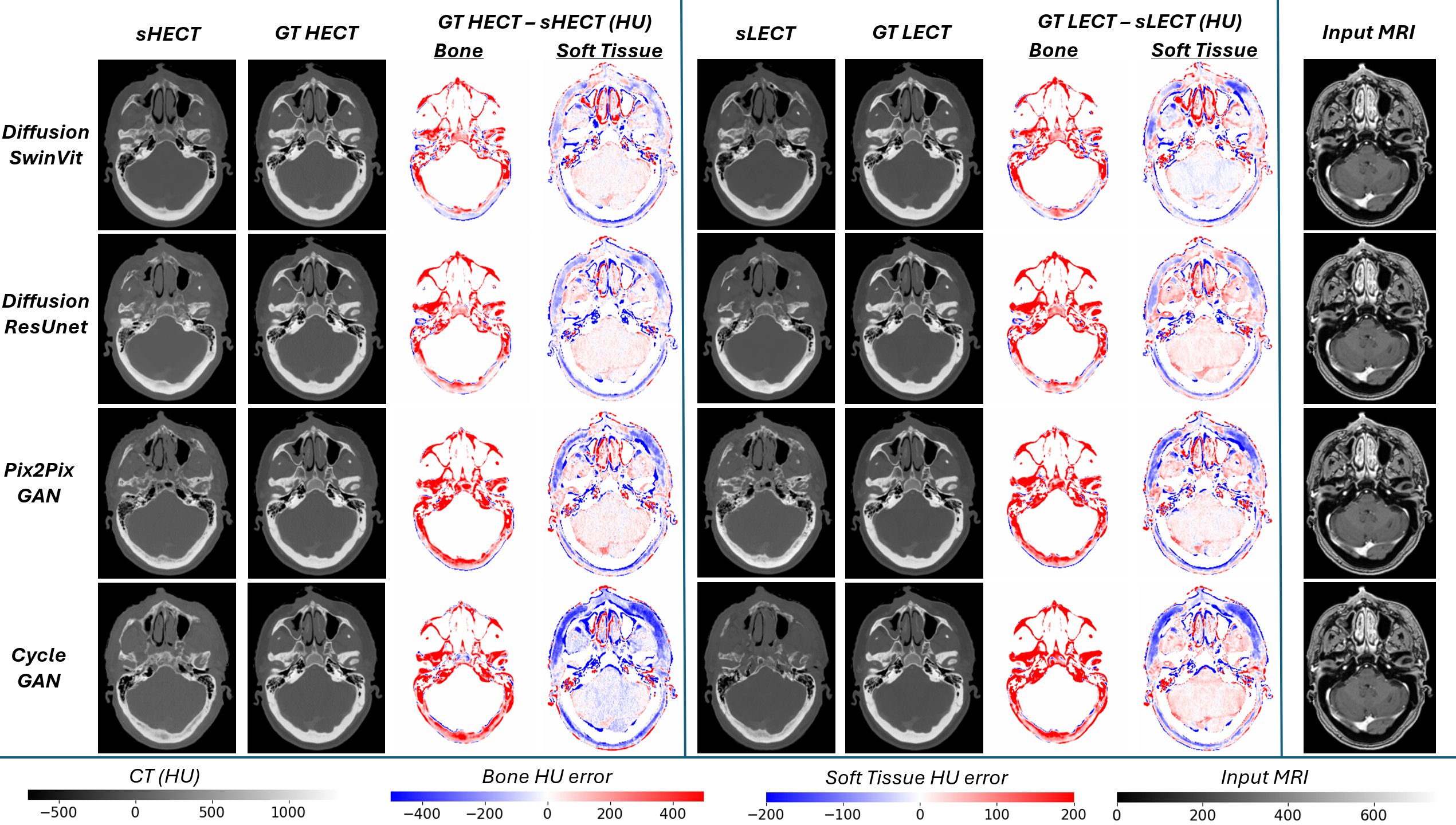}
    \caption{Representative axial slice of a test patient for comparing sHECT and sLECT images with respect to their GT counterparts. Our four DL-based models were assessed, from top to bottom rows: Diffusion-SwinVit, Diffusion-ResUNet, Pix2PixGAN, and CycleGAN. From left to right, the first two columns show sHECT and GT HECT, followed by two other columns with the corresponding HU error maps for bone and soft tissues. The same disposition is shown for sLECT and GT LECT in the next four right columns. The rightmost column displays the conditioning MRI input. HU error maps use a blue–red color scale to highlight underestimation (blue) versus overestimation (red) in HU.}
    \label{fig:HU_error_maps_1}
\end{figure*}

\begin{figure*}[!h]
    \centering
    \includegraphics[width=\textwidth]{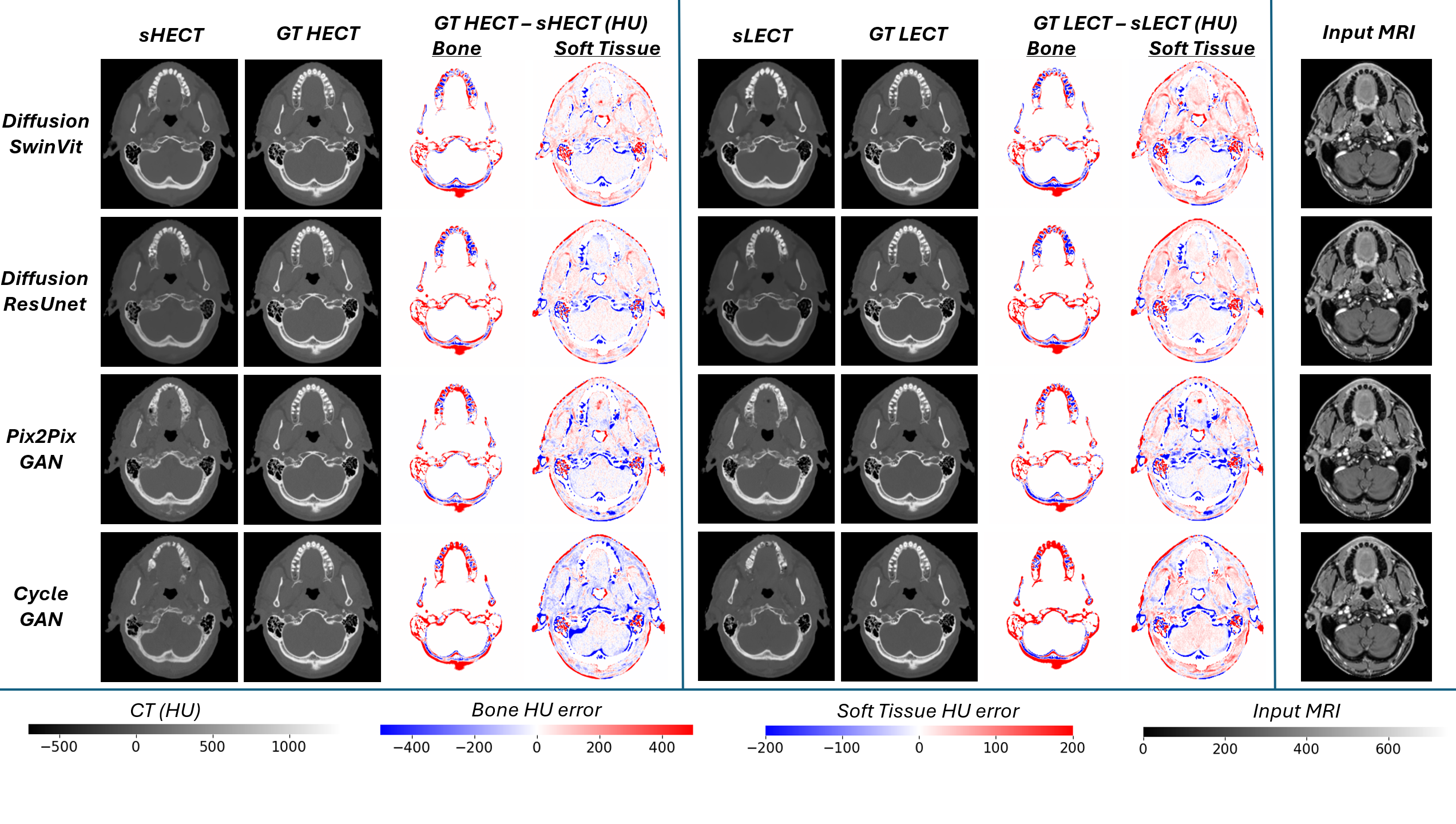}
    \caption{Representative axial slice of another test patient for comparing sHECT and sLECT images with respect to their GT counterparts. Our four DL-based models were assessed, from top to bottom rows: Diffusion-SwinVit, Diffusion-ResUNet, Pix2PixGAN, and CycleGAN. From left to right, the first two columns show sHECT and GT HECT, followed by two other columns with the corresponding HU error maps for bone and soft tissues. The same disposition is shown for sLECT and GT LECT in the next four right columns. The rightmost column displays the conditioning MRI input. HU error maps use a blue–red color scale to highlight underestimation (blue) versus overestimation (red) in HU.}
    \label{fig:HU_error_maps_2}
\end{figure*}

Synthetic images (sHECT, sLECT) were visually analyzed, together with HU difference maps for the bone and soft tissue regions for two different patients and two specific slices (Figures \ref{fig:HU_error_maps_1}-\ref{fig:HU_error_maps_2}). DDPM preserves detailed structures around the temporal bones, orbits, and mandibular region, resulting in fewer artifacts in areas that are typically difficult to reconstruct. This suggests that subtle morphological features, such as the bony contours near the sinuses or the intricate shape of the alveolar ridge around the teeth, are captured with higher fidelity. However, we note that the HU intensity values for the bone structures are all underestimated. The Diffusion-SwinVit model captures the bony anatomy more accurately, as evidenced in the HU difference maps by fewer and lower error intensity regions. The predicted HU values are in closer agreement with the GT volume, showing a more refined mapping within bone-soft tissue transitions. cGAN-based methods show a more diffuse and intense error pattern overall. Visually, this manifests as expansive patches of intense color in the difference maps of regions such as the skull base and mandible. Consequently, the boundary zones between bone and soft tissue sometimes lack enough definition for being used in clinical interpretation. Larger and more scattered underestimated regions of bone density error appear, implying higher local HU errors, particularly around complex bone structures such as the temporal bones and sinus cavities. 

Soft-tissue regions exhibit less pronounced discrepancies than in bone, particularly within the brain area, but they can still have clinical implications near critical structures or treatment margins. DDPM produces more homogeneous tissue intensities and natural transitions between soft tissues, where errors remain relatively localized and do not extend over large anatomical areas. They show fewer areas of mismatch and lower error magnitudes as opposed to cGAN-based methods, with more spatially confined deviations and a more faithful representation of tissue interfaces. Visual differences become especially apparent around boundaries between bone and soft tissue. Diffusion‐based models typically preserve the sharp transitions at the skull base and other complex bone‐tissue interfaces such as the teeth contours. Meanwhile, the GAN‐based predictions exhibit more blurring and textural artifacts, which can compromise the delineation of fine anatomical details. Moreover, the consistency in quality across the sHECT and sLECT scans appears to be stronger in the diffusion‐based approaches, suggesting that their learning mechanism captures a robust mapping between MRI features and the DECT outputs. In contrast, the cGAN‐based methods can exhibit noticeable variation in how well they reconstruct HECT versus LECT scans.  

\begin{figure*}[!h]
    \centering
    \includegraphics[width=\textwidth]{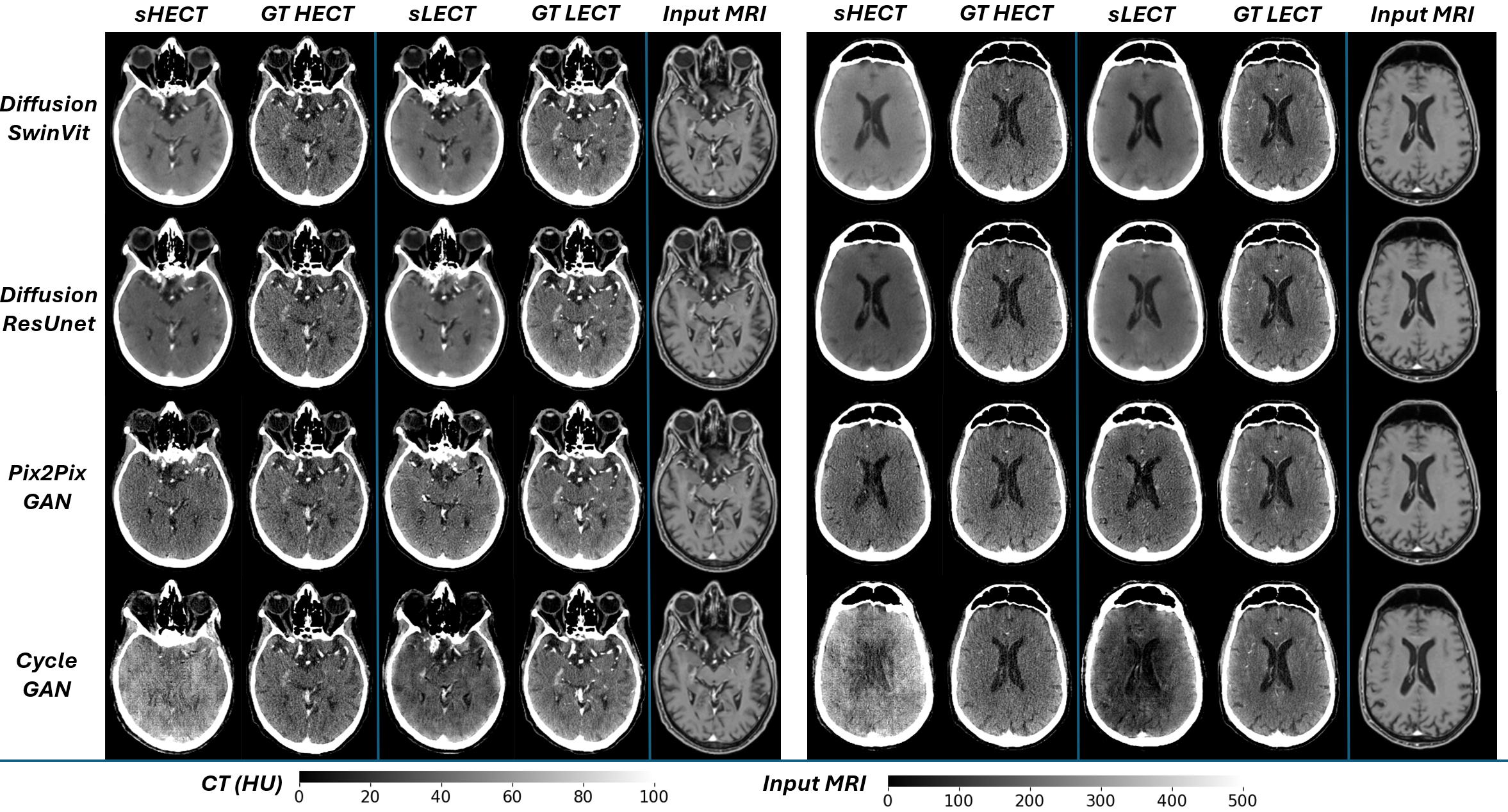}
    \caption{Qualitative comparison of brain ventricles and eyes regions for sHECT and sLECT images against their GT counterparts with regard to our four DL models (rows, from top to bottom): Diffusion-SwinVit, Diffusion-ResUNet, Pix2PixGAN, CycleGAN. For each of the two different axial slices, eyes (left group) and brain ventricles (right group), the first two left columns show the sHECT versus GT HECT comparison, the third and fourth columns show the sLECT vs GT LECT comparison, while the fifth column shows the MRI input. All images are displayed adjusting the window and level to emphasize soft tissue visualization.}
    \label{fig:qualitative_images}
\end{figure*}

Image quality and definition are assessed within soft tissues in Figure \ref{fig:qualitative_images}, particularly across key areas like the ventricles, the eyes, and the optic nerves. Visual inspection reveals that diffusion-based approaches produce smooth yet detailed soft-tissue regions. Inherent noise from GT images is removed but maintaining a closer resemblance to the original DECT images and keeping the consistency on HU values without abrupt shifts near high-contrast boundaries. Addressing noise and artifacts, the iterative denoising diffusion strategy is able to suppress fake patterns and artifacts, providing more uniform image quality across the entire volume. Diffusion-SwinVit demonstrates the most consistent image quality and definition among soft tissues. It preserves tissue boundaries particularly well, demonstrating very low noise, retaining subtle anatomical details, and preserving bone edges, while Diffusion-ResUNet images appear slightly smoother or less textured in some regions. On the other hand, cGAN-based models obscure fine distinctions between gray and white matter, suggesting a less stable intensity translation across the different tissue types and reducing their clinical reliability. Pix2PixGAN tends to have more contrast deviations and a higher noisy appearance than DDPM approach, introducing mild over/under emphasis of soft tissue intensities and textures. CycleGAN produces less defined textures and some blotchy regions with an increased heterogeneity in the brain parenchyma, also showing a chance of introducing patchy artifacts. This indicates mismatches in the learned intensity values mapping, which hampers the precise distinction of tissues. 

\begin{figure*}[!h]
    \centering
    \includegraphics[width=\textwidth]{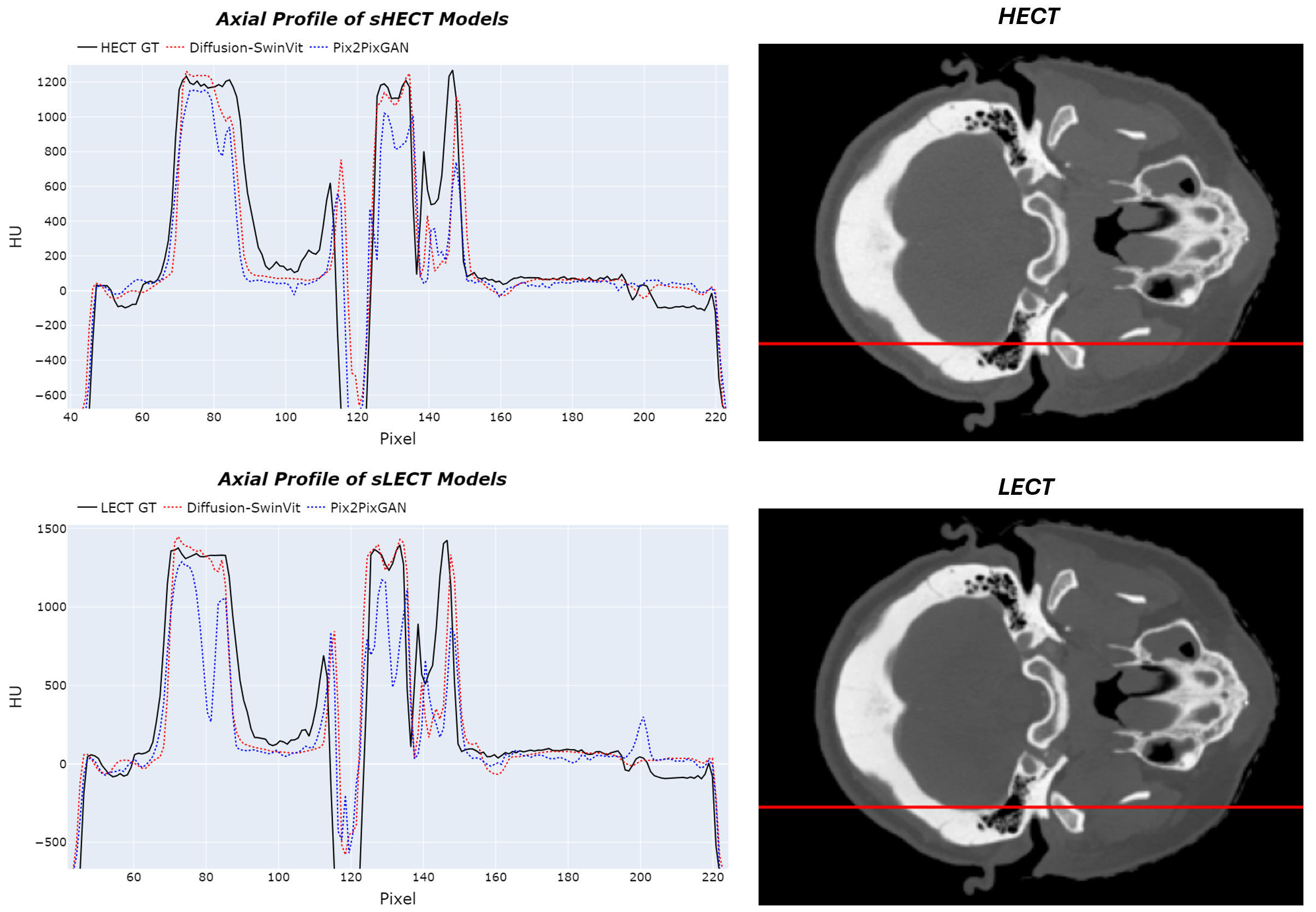}
    \caption{Axial profile of HU values for a specific patient slice. In the left plots, the solid black line represents the GT (HECT upper plot, LECT lower plot), while the red and blue dotted lines correspond to the respective synthetic images from Diffusion-SwinVit and Pix2PixGAN models. The image on the right of each plot identifies the location of the profile axial line (red horizontal line) across the slice, traversing bone and soft-tissue regions.}
    \label{fig:axial_profile}
\end{figure*}

Figure \ref{fig:axial_profile} showcases the axial profile evaluation. Diffusion-SwinVit model (red dotted line) tracks the GT more closely, particularly in areas of steep intensity changes such as the tissue boundary between bone and soft tissues. These transitions correspond to steep spikes in HU values (above 1000 HU), where accurate modeling of high-contrast structures is critical for subsequent clinical reliance. In contrast, Pix2PixGAN (blue dotted line) tends to deviate more substantially in these transition regions, often overestimating or underestimating HU values. Both profile plots highlight how the diffusion-based method remains relatively faithful to the sharp bone edges and avoids large spikes or drops in HU, whereas Pix2PixGAN shows bigger discrepancies around these peaks, indicating potential challenges in capturing rapid intensity transitions. This pattern is even more evident in LECT profile plot, where a notably clearer difference in performance between both models can be observed.

\begin{figure*}[!h]
    \centering
    \includegraphics[width=\textwidth]{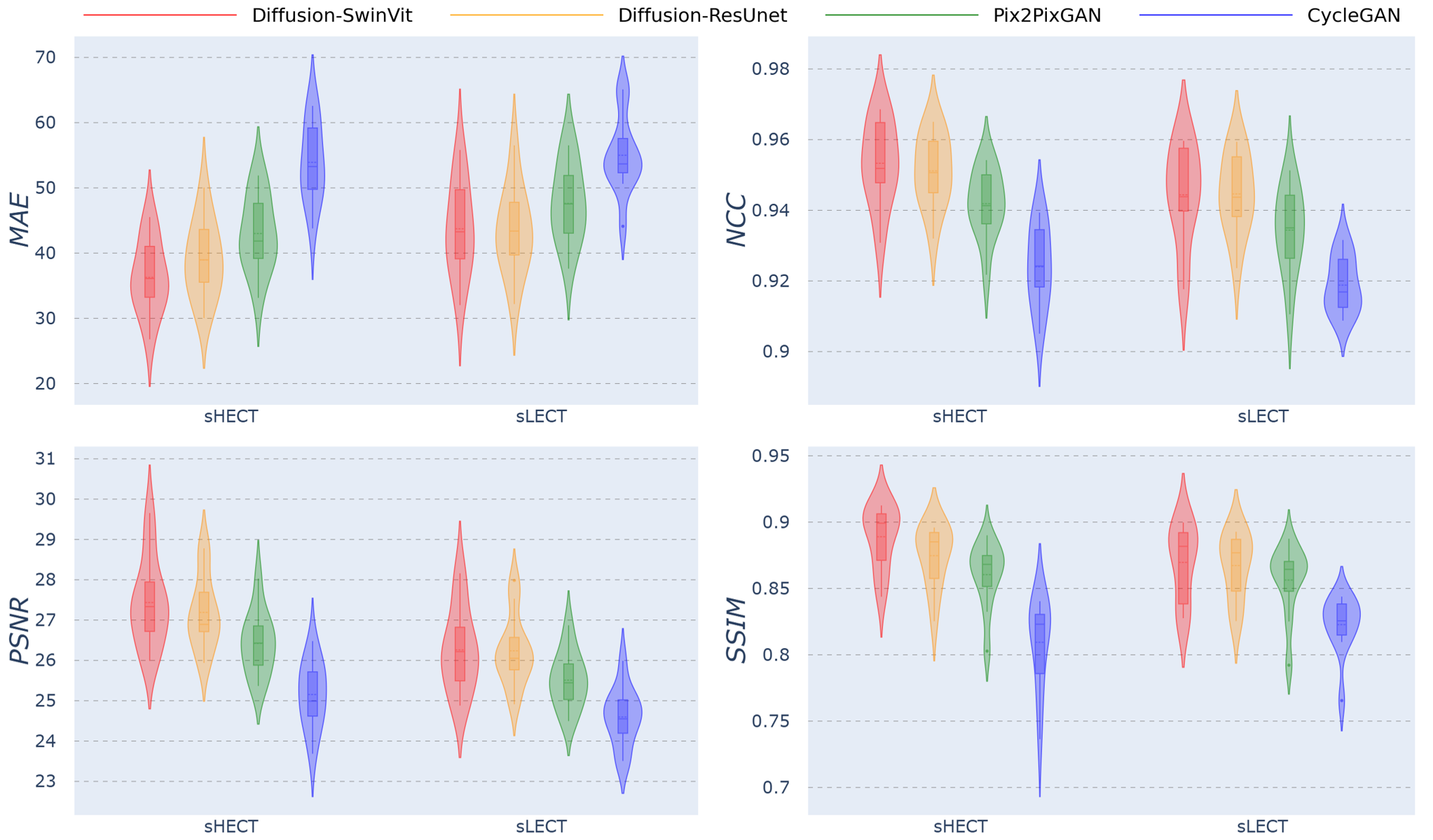}
    \caption{Violin plots for quantitative comparison of sHECT and sLECT images against their GT images among all test patients. Four key performance metrics (MAE, NCC, PSNR, and SSIM) were calculated across our four different models (Diffusion-SwinVit, Diffusion-ResUNet, Pix2PixGAN, and CycleGAN). Each violin plot is related to a model’s performance showing both the statistical distribution (width and shape) and a box-plot style summary (median, quartiles).}
    \label{fig:violin_plot_metrics}
\end{figure*}

Table \ref{tab:table_metrics} and Figure \ref{fig:violin_plot_metrics} both show the performance of the models regarding four image quality metrics: MAE, Peak Signal-to-Noise Ratio (PSNR), Normalized Cross-Correlation (NCC), and Structural Similarity Index Measure (SSIM). Overall, diffusion-based approaches consistently presents superior performance across all image metrics compared to cGAN-based models, as shown by the lower MAE, and higher NCC, PSNR, SSIM values achieved by our DDPM. This may indicate reduced reconstruction errors and better image similarity to the GT volumes. Pix2PixGAN, while performing better than CycleGAN, still shows an evident gap compared to diffusion-based methods. CycleGAN consistently underperforms the rest of the methods across all metrics and shows a broader distribution and variability. Diffusion-SwinVit shows slightly superior image metrics compared to Diffusion-ResUnet, suggesting an advantage of incorporating attention-based ST into the diffusion process. The same conclusions that have been drawn in the case of sHECT can be derived with sLECT models, despite the image metrics being slightly better for sHECT case. Statistical analysis revealed that the chosen model had a significant impact on all assessed metrics, as indicated by the ANOVA test ($p$-value $<$ $1\%$). Furthermore, subsequent Wilcoxon test was performed for every pair of models, and confirmed the presence of statistically significant differences among all models for each of the four metrics ($p's < 0.01$).

\begin{table}[h]
\centering
\caption{Comparative table showing four image metrics (MAE, PSNR, SSIM, NCC) computed between each synthetic and GT image. Mean and standard deviation were calculated over all test subjects.}
\label{tab:table_metrics}
\resizebox{\textwidth}{!}{
\begin{tabular}{lccccccc}
\toprule
\hline
Models & \textit{\textbf{MAE (HU)}} & \textit{\textbf{PSNR (dB)}}  & \textit{\textbf{SSIM (-)}} & \textit{\textbf{NCC (-)}}
        \\
        \hline
        \hline
        \textit{\textbf{Diffusion-SwinVit (HECT)}} 
        & 36.293 ± 5.250
        & 27.436 ± 1.057 
        & 0.889 ± 0.022 
        & 0.953 ± 0.011
        \\
        \hline
        
        \textit{\textbf{Diffusion-SwinVit (LECT)}} 
        & 43.721 ± 6.771
        & 26.259 ± 0.978 
        & 0.869 ± 0.026 
        & 0.944 ± 0.013
        \\
        \hline
        
        \textit{\textbf{Diffusion-ResUnet (HECT)}}
        & 38.955 ± 5.582
        & 27.191 ± 0.877  
        & 0.874 ± 0.021 
        & 0.951 ± 0.009
        \\
        \hline
        
        \textit{\textbf{Diffusion-ResUnet (LECT)}} 
        & 43.376 ± 6.372 
        & 26.237 ± 0.854 
        & 0.867 ± 0.023 
        & 0.944 ± 0.010
        \\
        \hline
        
        \textit{\textbf{Pix2PixGAN (HECT)}} 
        & 42.998 ± 5.421 
        & 26.420 ± 0.744 
        & 0.860 ± 0.023
        & 0.941 ± 0.008
        \\
        \hline
        
        \textit{\textbf{Pix2PixGAN (LECT)}} 
        & 47.626 ± 5.707  
        & 25.509 ± 0.667  
        & 0.856 ± 0.025  
        & 0.934 ± 0.011
        \\
        \hline
        
        \textit{\textbf{CycleGAN (HECT)}}  
        & 55.738 ± 9.381  
        & 25.153 ± 0.772 
        & 0.800 ± 0.049
        & 0.924 ± 0.010
        \\
        \hline

        \textit{\textbf{CycleGAN (LECT)}}  
        & 55.011 ± 5.740  
        & 24.595 ± 0.685 
        & 0.822 ± 0.021
        & 0.918 ± 0.007
        \\
        \hline
        
\hline
\bottomrule
\end{tabular}
}
\end{table}

\begin{figure*}[!h]
    \centering
    \includegraphics[width=\textwidth]{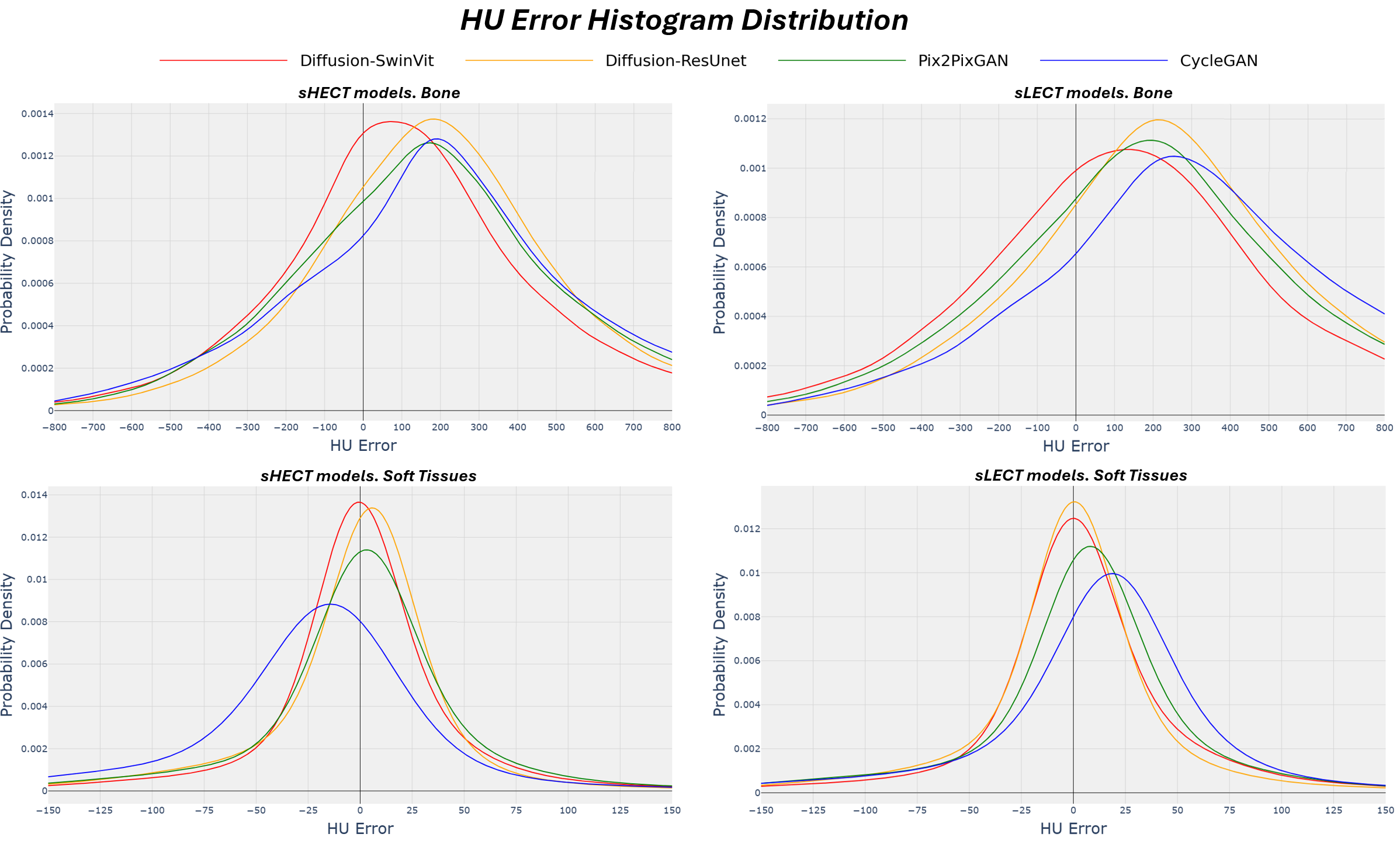}
    \caption{Comparative histogram distribution of HU prediction errors with regard to sHECT (left column) and sLECT (right column) scans along test subset. Results are particularly displayed for bone tissues (top row) and soft tissues (bottom row). Each colored curve represents a different model: Diffusion-SwinVit (red), Diffusion-ResUNet (orange), Pix2PixGAN (green), and CycleGAN (blue), respectively. The x-axis spans the HU error range (negative values indicate overestimation of GT, and positive values indicate underestimation), while the y-axis shows the probability density of HU error.}
    \label{fig:histogram_HU_error}
\end{figure*}

Figure \ref{fig:histogram_HU_error} provides an overall overview of HU error distribution plots. Diffusion-based methods appear to be the most robust as they produce narrower and more centered symmetric distributions around a zero error. This suggests better accuracy and consistency for HU predictions. The distribution peak is tighter and closer to zero, with the tails less pronounced, generating fewer large errors and outliers. On the contrary, cGAN-based models show broader curves, and they appear more shifted from zero error, indicating higher variability and a tendency to misestimate HU values. CycleGAN curve is the one more systematically skewed toward negative or positive HU errors. Error distributions within soft tissues are more concentrated around zero compared to bone distributions, as soft tissue HU lie in a narrower band of values, which makes their likely error range smaller. This reflects that high-density structures like bones pose a harder challenge for accurate synthesis since they require a precise modeling of large intensity jumps. 

\begin{figure*}[!h]
    \centering
    \includegraphics[width=\textwidth]{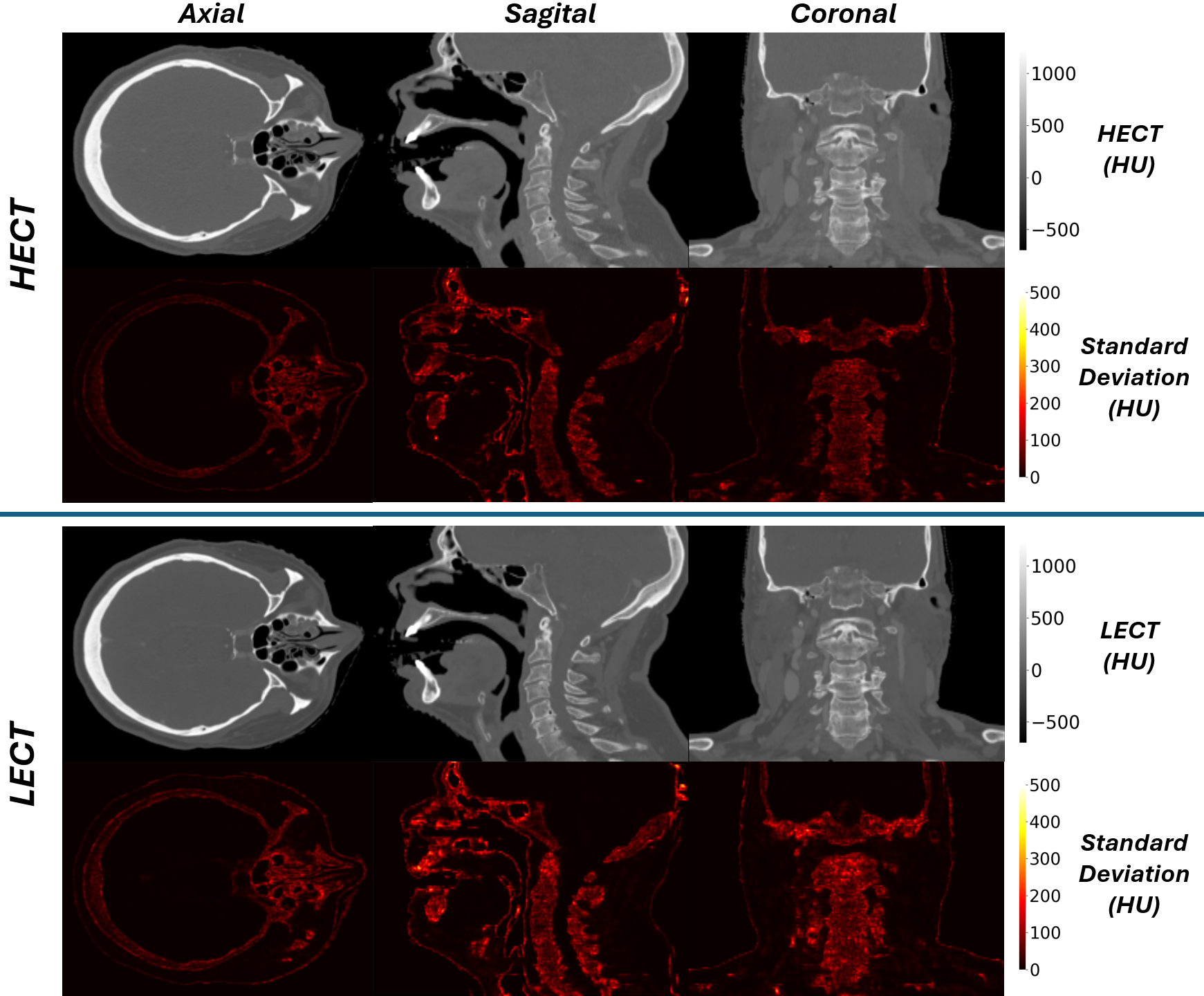}
    \caption{Uncertainty level, measured through standard deviation in absolute value, is highlighted for HU predictions across HECT and LECT scans regarding our DDPM-SwinVit approach. Five independent sDECT volumes were generated per test patient. The top section shows HECT slices in different planes (axial, sagittal, coronal) alongside their corresponding uncertainty heatmaps, while the bottom section provides the same for the LECT scan. The color scale ranges from black (near 0 HU, signifying low uncertainty) to bright red or yellow (up to 500 HU, indicating higher uncertainty).}
    \label{fig:uncertainty_maps}
\end{figure*}

We also evaluated the uncertainty of sDECT volumes generated with our DDPM approach (Figure \ref{fig:uncertainty_maps}). Voxel-wise standard deviation maps were derived from five independently generated sDECT volumes from each of the test patients. Visually, much of the brain parenchyma and other relatively uniform soft-tissue regions appear in darker tones, reflecting low dispersion across the multiple samples and a higher model confidence. In contrast, intense red color are prone to concentrate around interfaces between distinct tissue regions and within bone areas such as the skull, nasal sinuses, and vertebrae in the sagittal slices, indicating that the model is more prone to variability in areas of abrupt HU changes or complex anatomical structures like the mandible and upper spine. Comparing the high-energy versus low-energy maps, although both share a similar distribution pattern of uncertain voxels, the level of uncertainty becomes slightly higher for complex bony regions in sLECT volume. This is consistent with the qualitative and quantitative results shown in this section, where sHECT performance is somewhat superior compared to sLECT. 

{To assess the applicability of our approach into a realistic PT scenario, a SPR analysis was performed. SPR maps were generated from sDECT data and subsequently compared with those obtained from GT DECT acquisitions. Rather than targeting an improvement over real DECT performance, the aim was to demonstrate that sDECT can reproduce SPR values with sufficient consistency to be considered clinically usable.}

\begin{figure*}[!h]
    \centering
    \includegraphics[width=\textwidth]{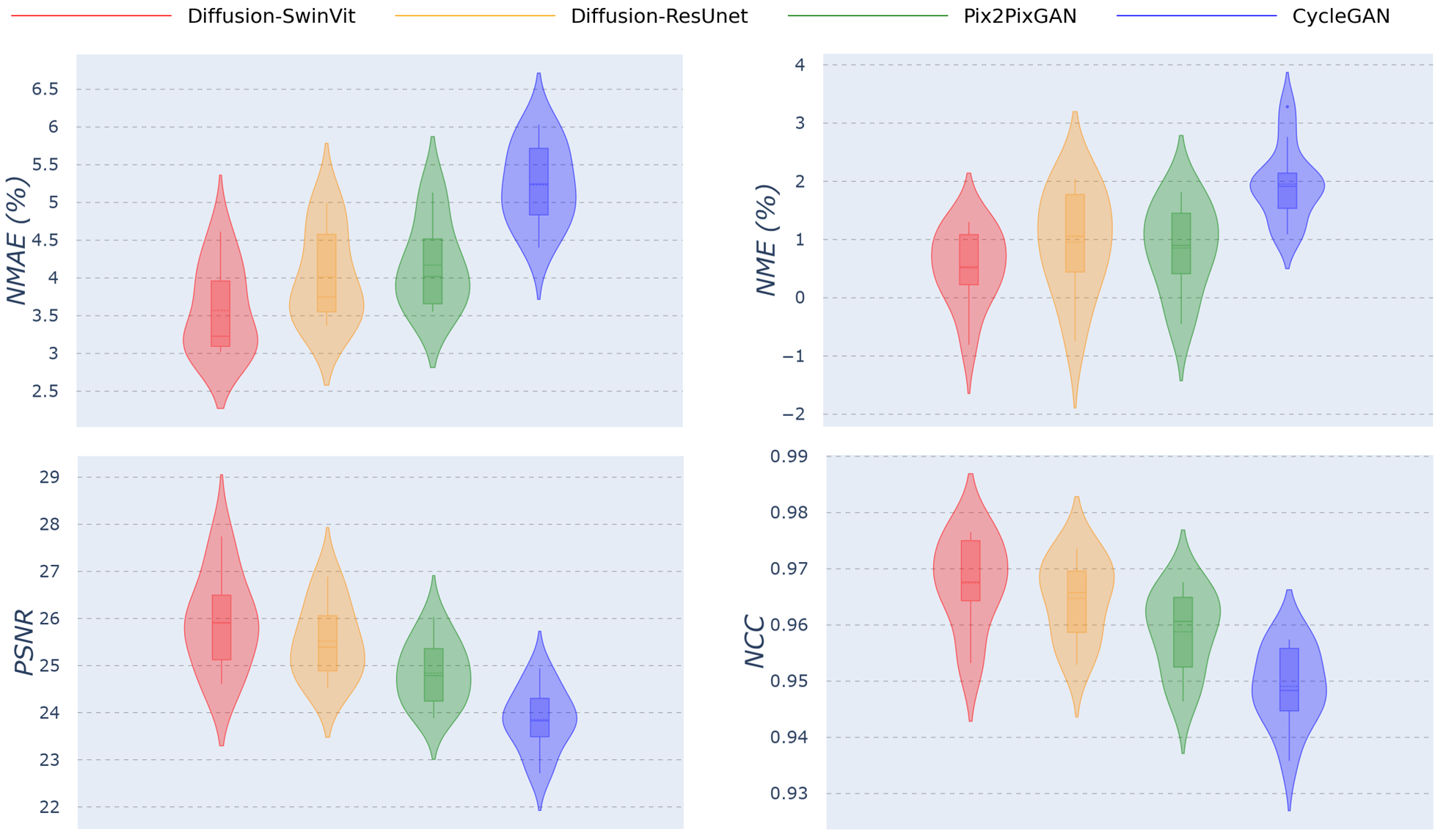}
    \caption{{Violin plots for the quantitative comparison between SPR values obtained from sDECT and SPR values derived from GT DECT across all test patients. Four key performance metrics (NMAE, NME, PSNR, and NCC) were calculated across four different models (Diffusion-SwinVit, Diffusion-ResUNet, Pix2PixGAN, and CycleGAN). Each violin plot shows both the statistical distribution (width and shape) and a box-plot style summary (median, quartiles) of a model in particular.}}
    \label{fig:violin_plot_SPR}
\end{figure*}

{A consistent SPR performance hierarchy emerges across both the numerical metrics summary and the distributional population analysis. Table \ref{tab:table_SPR} highlights the differences in terms of mean performance, while the violin plots of Figure \ref{fig:violin_plot_SPR} reveal how these advantages persist across the entire cohort of patients. The diffusion-based models, in particular the Diffusion-SwinViT architecture, provide more accurate SPR estimates across patients by systematically achieving lower NMAE and NME while higher PSNR and NCC compared with the cGAN-based approaches. Comparing architectures, Diffusion-SwinViT outperforms Diffusion-ResUNet and Pix2PixGAN outperforms CycleGAN.}

\begin{table}[h]
\centering
\caption{{Comparative table showing four quantitative error metrics (NMAE, NME, PSNR, NCC) computed between SPR values obtained from sDECT and GT DECT respectively. Mean and standard deviation were calculated over all test subjects and shown for each metric and model.}}
\label{tab:table_SPR}
\resizebox{\textwidth}{!}{
\begin{tabular}{lccccccc}
\toprule
\hline
Models & \textit{\textbf{NMAE (\%)}} & \textit{\textbf{NME (\%)}}  & \textit{\textbf{PSNR (dB)}} & \textit{\textbf{NCC (-)}}
        \\
        \hline
        \hline
        \textit{\textbf{Diffusion-SwinVit}} 
        & 3.569\% ± 0.544 \%
        & 0.532\% ± 0.611 \% 
        & 25.918 ± 0.951 
        & 0.967 ± 0.007
        \\
        \hline
        
        \textit{\textbf{Diffusion-ResUnet}}
        & 4.015\% ± 0.572 \%
        & 0.955\% ± 0.838 \%  
        & 25.519 ± 0.760 
        & 0.965 ± 0.007
        \\
        \hline
        
        \textit{\textbf{Pix2PixGAN}} 
        & 4.170\% ± 0.540 \%  
        & 0.855\% ± 0.710 \%  
        & 24.842 ± 0.637  
        & 0.959 ± 0.007
        \\
        \hline
        
        \textit{\textbf{CycleGAN}}  
        & 5.233\% ± 0.497 \%  
        & 1.601\% ± 1.276 \% 
        & 23.853 ± 0.629
        & 0.949 ± 0.006
        \\
        \hline
        
\hline
\bottomrule
\end{tabular}
}
\end{table}

\begin{figure*}[!h]
    \centering
    \includegraphics[width=\textwidth]{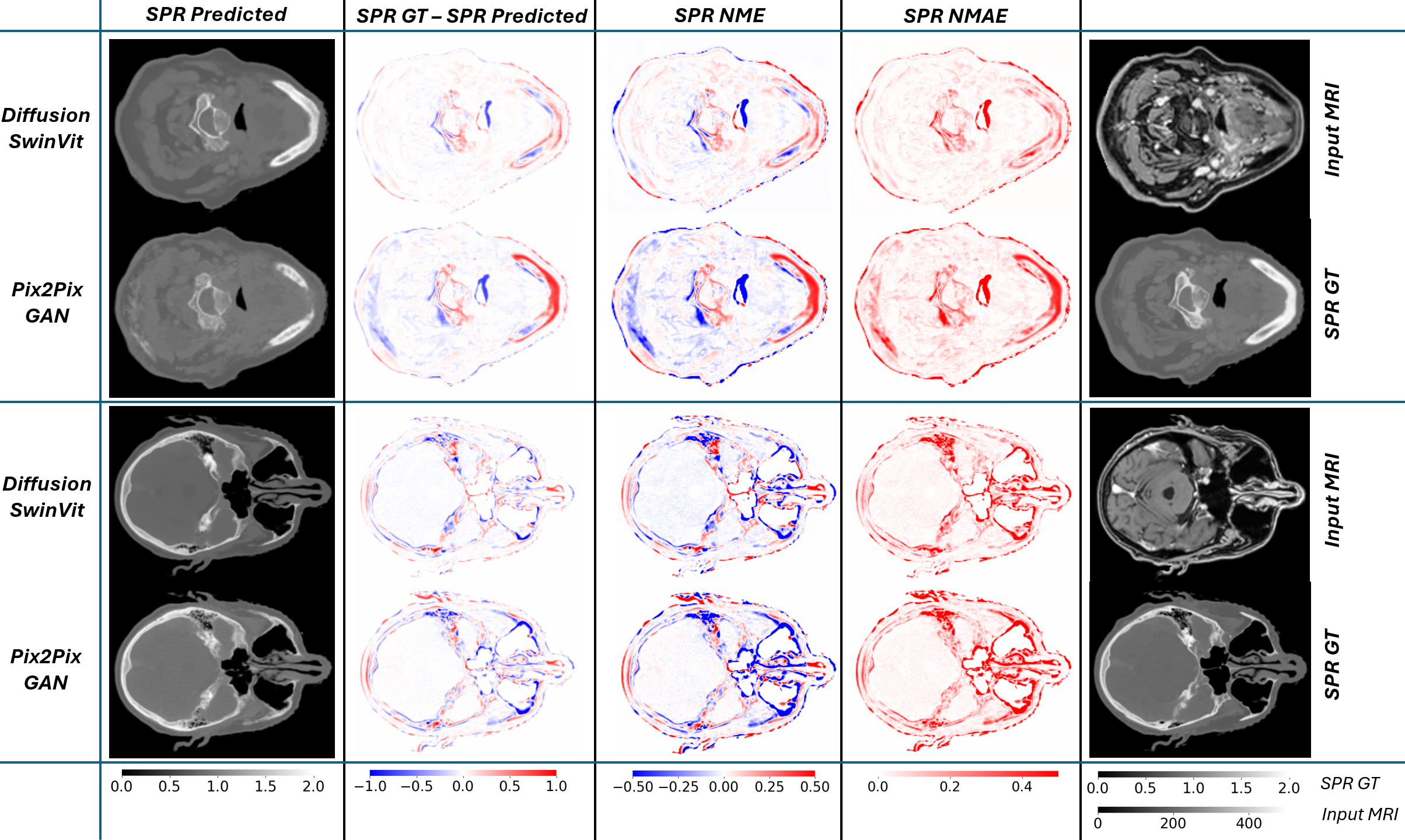}
    \caption{{Comparison of SPR maps derived from sDECT and their GT DECT counterparts from two different axial slices. The first and second rows correspond to one test patient, while the third and fourth rows correspond to another test patient. Diffusion-SwinVit and Pix2PixGAN models were assessed. From left to right: the predicted SPR map, the SPR map difference, NMAE SPR error map and NME SPR error map. The rightmost column displays the MRI input and the GT SPR map.}}
    \label{fig:SPR_error_maps}
\end{figure*}

{For the SPR error-map analysis, we compared the strongest diffusion-based model (Diffusion-SwinViT) with the leading cGAN variant (Pix2PixGAN) on two representative test patients. The qualitative and voxel-wise error analyses shown in Figure \ref{fig:SPR_error_maps} reveal a consistent pattern across the different anatomical regions. Diffusion-SwinViT model maintains lower error magnitudes in the spatial error distribution. NME, NMAE and SPR difference maps illustrate how the deviations of Diffusion-SwinViT are concentrated predominantly along complex anatomical interfaces such as bone-soft tissue boundaries while preserving the expected SPR contrast within homogeneous regions, indicating fewer systematic over/underestimations of SPR values across different tissues. However, the error maps of Pix2PixGAN show more pronounced blue/red error regions and broader regions of deviation, suggesting lower SPR stability across tissues.}

{When considered together, Table \ref{tab:table_SPR}, Figure \ref{fig:violin_plot_SPR}, and Figure \ref{fig:SPR_error_maps} provide a coherent multi-level validation of the estimated SPR. DDPM framework shows not only better SPR average performance but also yields improved robustness at the population level and anatomically faithful distribution of SPR estimates. The combination of lower errors, higher fidelity, and greater robustness across individuals reinforces the suitability of using diffusion-based sDECT images for SPR calculation.}

\section{Discussion}

This study proposes a conditional DDPM–based image-to-image framework for synthesizing DECT images from MRI data, marking a pivotal step toward enhancing image-guided RT and PT. Recent developments in synthetic medical imaging have enabled the generation of DECT-like information directly from MRI, leveraging MRI’s superior soft-tissue contrast and eliminating ionizing radiation exposure, which is especially harmful in pediatric patients and associated with numerous adverse effects. DECT enables more accurate tissue characterization and SPR estimation compared to SECT scans, reducing range uncertainties and increasing the overall precision of PT treatment. Since it is sometimes not feasible to increase the radiation margins due to the proximity of critical OAR structures, our framework makes it more likely for clinicians to maintain treatment precision without compromising patient safety. Given that paired MRI-DECT scans are obtained on the same day, inter-modality variability is minimized and the anatomical alignment is expected to be highly consistent. The performance of the proposed DDPM is demonstrated through qualitative and quantitative analysis of sDECT images: image quality metrics, HU line profiles, HU error maps, histograms of HU error distribution with regard to bone and soft tissues regions, uncertainty level maps, SPR error metrics, SPR error maps. Our DDPM notably enhances sDECT image quality compared to cGAN approaches (Pix2PixGAN, CycleGAN).

Although DL-based approaches were previously explored to synthesize DECT images from a range of input modalities (SECT, CBCT, MRI), diffusion-based models remained unexplored for MRI-based sDECT image generation, making this study the first to investigate this framework. While cGANs have demonstrated success in MR-to-CT translation tasks (see Related Work section), they are still prone to a range of training issues (instabilities, hyperparameter sensitivities, mode collapse) that often undermine their reliability and produce unrealistic or inconsistent image outputs. DDPM solves many of these cGAN drawbacks by stabilizing the training process, reducing the computational overhead necessary for hyperparameter tuning, and avoiding mode collapse. Hence, DDPM have emerged as a superior alternative compared to cGANs, due to their rigorous theoretical foundation, smoother training process, and ability to produce high-quality images, making them particularly well-suited for medical imaging synthesis applications. 

Several factors account for Pix2Pix’s superior performance over CycleGAN. Since our training data was previously registered, Pix2PixGAN combined its adversarial objective with a voxel-wise L1 reconstruction loss. At every gradient update the generator is explicitly punished for HU discrepancies at the exact locations of bone edges, air cavities and soft tissues. sDECT images are not just plausible to a discriminator but numerically faithful to the GT on a per-voxel basis. CycleGAN, by design, dispenses with paired supervision and instead learns two adversarial mappings bounded only by cycle-consistency loss, which allows the generated DECT images to be anatomically mis-aligned or intensity-shifted, so long as it can still be transformed back into the original MRI. As well, CycleGAN deploys two generator–discriminator pairs, effectively doubling the adversarial losses that must be balanced. This added complexity makes parameters optimization less stable, provoking more training oscillations and slower convergence, often steering the model toward inferior local minima.

LECT acquisitions (lower kVp) provide high contrast at the expense of increasing the image noise. Lower-energy photons are more easily absorbed or scattered by the body, meaning fewer and more randomized photons reach the detector. This results in a weaker signal, stronger noise, and a reduced signal-to-noise ratio (SNR). In contrast, HECT scans (higher kVp) exhibit lower contrast but benefit from reduced noise and improved spatial resolution. Higher-energy photons penetrate tissues more effectively and are less likely to be absorbed or scattered, allowing more photons to reach the detector and producing more consistent detector signals. Thus, they benefit from better SNR because the image is formed with a larger and more consistent photon flux \cite{borges2023pros}. These physical properties influenced the performance of our synthesis methods. The superior performance obtained with our DL models in synthesizing HECT images compared to LECT can be attributed to the higher SNR and more stable anatomical detail of HECT scans, which provide a more consistent mapping target during training. The elevated noise and contrast variability in LECT images make their image synthesis more challenging, especially when the input is derived from MRI, which does not inherently reflect CT contrast dynamics.  

{MRI volumes were deformed and resampled to match DECT geometry, while the original DECT images were left unaltered because they needed to be the registration reference. If the structural discrepancies between sDECT and GT DECT images were originated from registration inaccuracies, the magnitude and distribution of image error would appear consistently across models and the difference of model performance would not have a strong impact on the overall voxel-wise difference. However, as shown along the HU error maps (Figures 3,4) and axial profiles (Figure 6), the distribution and magnitude of the HU error vary substantially across models. Diffusion-based methods produce markedly reduced errors and smoother error patterns compared with cGAN-based methods, which would not be the case if the discrepancies were driven mostly by misregistration. The mismatches highlighted in the line profiles occur primarily at high-gradient anatomical boundaries, which are well known to be highly sensitive to residual spatial misalignments. Even sub-voxel differences between modalities can produce apparent HU shifts in line plots. Since MRI does not provide direct information about tissue density, bony regions and sharp anatomical transitions in sDECT images are expected to have larger uncertainty and HU discrepancies. This explains why all models exhibit more pronounced deviations in bone or bone-soft tissue interfaces, whereas soft-tissue regions (where MRI signal is more informative) show smaller and more homogeneous errors.}

{The quality and accuracy of sDECT images have been evaluated across previous works. Our DDPM shows noticeable quantitative and qualitative improvements in sDECT compared to MR-only label-GAN results we previously obtained in \cite{liu2021synthetic}. The quantitative and qualitative image results of our DDPM are also aligned to those reported by \cite{charyyev2022learning} and \cite{gao2024ct}, who performed DL-based DECT synthesis from SECT. In our case, the task is even more challenging because the input modality is MRI and we lack direct information about tissue density. Besides exploring the potential of our DDPM in MRI-based DECT synthesis, we performed a SPR evaluation study to assess the clinical reliability and applicability of sDECT images. The SPR maps derived from sDECT generated with our Diffusion-SwinVit method outperformed those derived using our best cGAN, and showed only minimal differences compared to SPR maps obtained from GT DECT, while the SPR uncertainty and the magnitude of SPR error metrics stayed within the expected acceptable range. Comparing the results of different methods, we demonstrate that greater sDECT image quality and accuracy translates into more precise SPR maps.}

Despite the strong potential of our proposed DDPM, certain constraints must be acknowledged. One major barrier to the clinical adoption of DDPM is the time it takes to generate images during inference, as it becomes a critical factor in time-sensitive environments such as medical imaging. Generating a single volume with our DDPM takes around 240 seconds, considerably slower than the approximate 10 seconds needed when using cGAN-based alternatives like Pix2Pix or CycleGAN. This performance time gap is primarily due to the iterative sampling process inherent to DDPM. Unlike GANs, which generate outputs in a single forward pass, diffusion models involve the sequential application of the generative network across multiple timesteps to perform a gradual denoising approach. This results in sampling times that scale linearly with the number of denoising timesteps, leading to significantly longer inference durations. Therefore, reducing sampling time without compromising image fidelity represents an essential future work for translating the advantages of diffusion models into real-world medical applications. Future studies should also handle the underlying variability of sDECT images, due to the application of random Gaussian noise along the forward diffusion phase.

{While the quantitative improvements observed for the diffusion-based models over adversarial approaches are modest in conventional image-based metrics, this behavior is not unexpected given the already high level of performance achieved by modern image-to-image translation frameworks. In this context, the advantage of diffusion-based models should be interpreted not only in terms of aggregate metric differences, but also in terms of their training stability and consistency in image synthesis. Unlike adversarial methods, which may be susceptible to training instability and potential mode collapse, diffusion probabilistic models are formulated through a well-defined stochastic forward–reverse process that promotes stable optimization and reproducible convergence~\cite{chang2024high}. This property is particularly relevant for medical imaging applications, where robustness across patients is essential.}

{More importantly, the clinical relevance of the proposed framework is closely linked to the role of image-derived quantities in proton therapy. Proton dose calculation is fundamentally governed by material-dependent stopping power, and accurate estimation of relative stopping power (RSP) or stopping power ratio (SPR) is a key determinant of proton range and dose accuracy. Prior studies have shown that uncertainties in CT-based material characterization represent a major source of proton range uncertainty, and that improvements in material estimation can translate into improved agreement in dose and water-equivalent thickness (WET) distributions~\cite{paganetti2012range,chang2023multimodal}. In addition, standardized Monte Carlo commissioning frameworks have further emphasized that accurate RSP modeling directly impacts proton transport, range prediction, and heterogeneous dose calculation~\cite{chang2020standardized}. Consistent with this physical relationship, previous validation studies demonstrated that improved deep learning-based material estimation leads to closer agreement between measured and simulated proton dose and range metrics~\cite{chang2022dual,Chang_2022_validation_dl}. Ultimately, these considerations suggest that even modest improvements in image-based performance may be relevant if they contribute to more reliable and physically consistent SPR estimation, which is closely related to proton dose and range calculation. However, the extent to which these improvements translate into measurable gains in dose or range accuracy was not explicitly evaluated in the present study and remains a limitation. Future work incorporating dedicated dose and proton range validation will be important to further assess the clinical impact of the proposed approach. In this context, the stable and consistent image synthesis behavior of the diffusion-based framework may provide a useful basis for such investigations and potential integration into proton therapy workflows.}

Our dataset comprised images from 53 patients. While enough to evaluate the efficacy of our approach, our model will likely see a reduction in performance in a multi-institutional dataset due to differing imaging acquisition parameters and patient demographics. Gathering a large-scale multimodal dataset remains inherently challenging, limiting both the variety and volume of data available for training. Therefore, expanding the dataset to include a more extensive and multi-center cohort would probably increase the robustness and reliability of the resulting model. Additionally, exploring different MRI modalities could help confirm that sDECT performance is not dependent on a single MRI sequence. A further refinement would be to harmonize MRI acquisition protocols, specifically matching the field of view (FOV) and slice counts, to align closely with those used for DECT scans, thus avoiding some preprocessing stages and ensuring a more direct and consistent correlation between imaging modalities. 
Since our model was trained on only head and neck images, extending our approach to additional anatomical regions would be a logical next step. 

There also exist some concerns regarding MRI and DECT image modalities. MRI inhomogeneity may undermine accurate segmentation and compromise the reliability of dose calculations based on SPR. Similarly, MRI can struggle to capture adequate compositional detail in the presence of heavy calcifications, severe osteoporosis, or metallic implants. Since we purely address on an image-to-image translation paradigm where fundamental principles of the imaging acquisition process are not explicitly captured, our approach may potentially produce outputs that are visually convincing yet physically inaccurate. DECT artifacts such as scatter or metal distortions might also introduce unwanted noise affecting model training. Furthermore, even with deformable registration, there could still be slight misalignments between DECT and MRI scans, impacting the evaluation of sDECT images. 

Looking ahead, future work should focus on DL–based solutions for improved inhomogeneity correction in MRI, artifact reduction in DECT, and more robust inter-modality registration methods. We should also incorporate physics principles to improve both the realism and clinical relevance of sDECT images and avoid relying only on the image-to-image conversion pipeline. All these approaches would help mitigate the possible sources of error mentioned, ensuring more reliable sDECT images. 

\section{Conclusion}

{Ongoing developments in radiation oncology call for increasingly advanced imaging techniques to meet the growing complexity of patient treatment. In this work, we propose a novel DL data-driven strategy that leverages the strengths of DDPM to generate sDECT images directly from MRI scans, demonstrating effectiveness to model the complex relationship between MRI and DECT modalities for simultaneously generating two-energy-level CT images (LECT and HECT), and outperforming established cGAN-based approaches. Our proposed DDPM produces accurate sDECT images with high anatomical fidelity, and the results of SPR estimation reveal a promise for their applicability along PT clinical workflow. Our goal is to further extend the clinical potential of DECT in PT through MRI-only workflows, avoiding patient exposure to harmful ionizing radiation. We aim not only to optimize the imaging pipeline workflow but also to support the broader transition toward MRI-only planning approaches, ultimately enabling more individualized and adaptive treatment strategies for oncology patients.}

\section*{{Data availability statement}}
{The data cannot be made publicly available upon publication because they are owned by a third party and the terms of use prevent public distribution. The data that support the findings of this study are available upon reasonable request from the authors.\\}






\section*{References}
\addcontentsline{toc}{section}{\numberline{}References}





\bibliography{refs.bib}      



\bibliographystyle{dcu}   


\end{document}